\documentclass{aa}
\usepackage{amsmath,amssymb}
\usepackage{graphicx}
\usepackage{txfonts}
\usepackage{natbib}
\bibpunct{(}{)}{;}{a}{}{,} 

\newcommand{\dd}{{\bf d}}
\newcommand{\kk}{{\bf k}}
\newcommand{\yy}{{\bf y}}

\newcommand{\cM}{{\cal M}}
\newcommand{\RR}{{\mathbf R}}

\newcommand{\YY}{{\mathbf Y}}

\newcommand{\cR}{{\cal R}}

\newcommand{\1}{{\bf 1}}

\begin{document}

\title{Filaments in observed and mock galaxy catalogues}
\author{R. S. Stoica\inst{1} 
	\and V. J. Mart{\'\i}nez\inst{2} 
	\and E. Saar\inst{3}
}
\institute{Universit\'e Lille 1, Laboratoire Paul Painlev\'e,
59655 Villeneuve d'Ascq Cedex, France\\
	\email{radu.stoica@math.univ-lille1.fr}
  \and Observatori Astron\`omic and Departament d'Astronomia i Astrof{\'\i}sica, Universitat de Val\`encia, Apartat de correus 22085, E-46071 Val\`encia, Spain\\
	\email{martinez@uv.es}
  \and Tartu Observatoorium, T\~oravere, 61602 Estonia\\
 	\email{saar@aai.ee}
}
\date{Received  / Accepted }

\abstract
{The main feature of the spatial large-scale galaxy distribution is an
intricate network of galaxy filaments. Although many attempts have been made
to quantify this network, there is no unique and satisfactory recipe for that yet.
}
{The present paper compares the filaments in the real data and in the numerical models, to see if our best models reproduce statistically the filamentary network of galaxies.
}
{We apply an object point process with interactions (the Bisous process) to trace
and describe the filamentary network both in the observed samples (the 2dFGRS catalogue) and in the numerical models that have been prepared to mimic the data. We compare
the networks.
}
{
We find that the properties of filaments in numerical models (mock samples) have a large variance. A few mock samples display filaments that resemble the observed filaments, but usually the model filaments are much shorter and do not form an extended network.
}
{
We conclude that although we can build numerical models that are similar to observations in many respects, they may fail yet to explain the filamentary structure seen in the data.
The Bisous-built filaments are a good test for such a structure.
}


\maketitle

\section{Introduction}

\begin{figure*} 
\centering 
\resizebox{0.8\textwidth}{!}{\includegraphics*{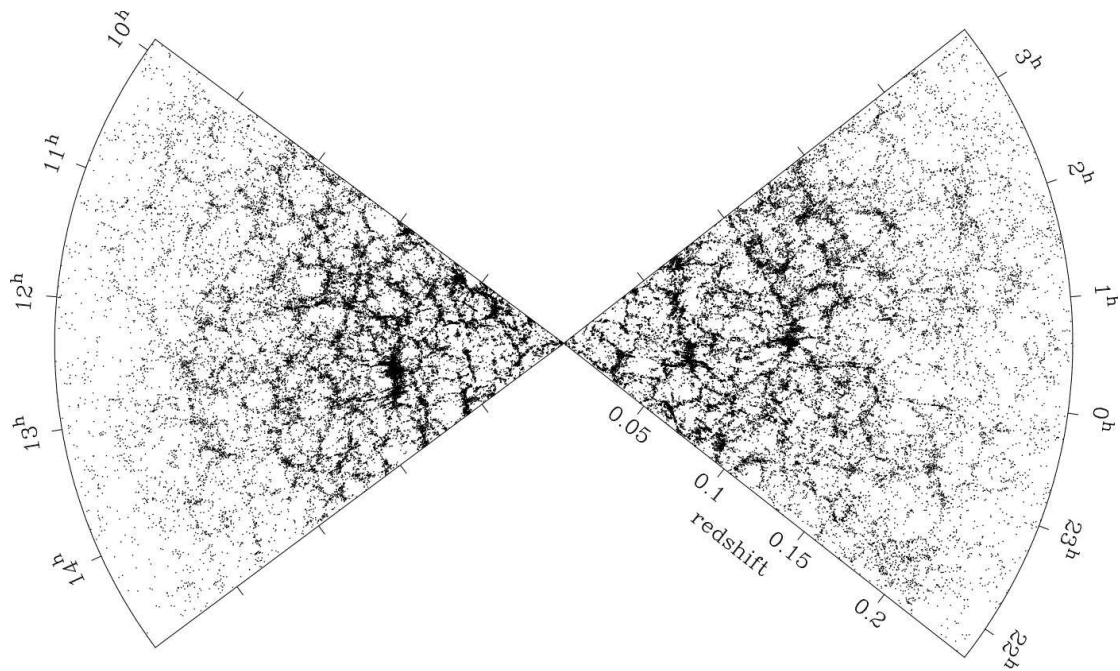}}\\ 
\caption{Galaxy map for two 2dFGRS slices of the
thickness of $2.6^\circ$.  The filamentary network of galaxies is 
clearly seen.
\label{fig:2dfgrs} } 
\end{figure*}

The large-scale structure of the Universe traced by the
three-dimensional distribution of galaxies shows intriguing patterns:
filamentary structures connecting huge clusters surround nearly empty
regions, the so-called voids. As an example, we show here a map from
the 2dF Galaxy Redshift Survey (2dFGRS, \citet{2dFGRS}). As an illustration
of the filamentary network, Fig.~\ref{fig:2dfgrs} shows
the positions of galaxies in two $2.6^\circ$ thick slices from two
spatial regions that the 2dFGRS covered. Distances are given in
redshifts $z$.

Filaments visually dominate the galaxy maps. Real three-dimensional
filaments have been extracted from the galaxy distribution as a result
of special observational projects \citep{pimbblet04a}, or by searching
for filaments in the 2dFGRS catalogue \citep{pimbblet04b}. These
filaments have been searched for between galaxy clusters, determining
the density distribution and deciding if it is filamentary,
individually for every filament \citep{Pimbblet05}.  Filaments are
also suspected to hide half of the warm gas in the Universe; an
example of a discovery of such gas is the paper by \citet{Werner2008}.

However, there are still no standard methods to describe the observed
filamentary structure, but much work is being done in this direction.
The usual second-order summary statistics as the two-point correlation
function or the power spectrum do not provide morphological
information. Minkowski functionals, minimal spanning tree (MST),
percolation and shapefinders have been introduced for this purpose
(for a review see \citet{martsaar02}).
 
The minimal spanning tree was introduced in cosmology by
\citet{barrow85}. It is a unique graph that connects all points of the
process without closed loops, but it describes mainly the local
nearest-neighbour distribution and does not give us the global and
large-scale properties of the filamentary network. A recent
development of these ideas is presented by \citet{Colberg07}. He
applies a minimal spanning tree on a grid, and works close to the
percolation regime -- this allows the study of the global structure of the
galaxy distribution. We note that using a grid introduces a smoothed
density, and this is typical for other recent approaches, too.

In order to describe the filamentary structure of continuous density
fields, a {\it skeleton} method has been proposed and developed by
\citet{eriksen04} and \citet{novikov06}.  The skeleton is determined
by segments parallel to the gradient of the field, connecting saddle
points to local maxima. Calculating the skeleton involves
interpolation and smoothing the point distribution, which introduces
an extra parameter, which is the band-width of the kernel function used to
estimate the density field from the point distribution, typically a
Gaussian function. This is generally the case for most of the density-based
approaches.  The skeleton method was first applied for two-dimensional
maps, an approach to study the cosmic microwave sky background
\citep{eriksen04}. The method was adapted for 3-D maps
\citep{Sousbie2008a} and was applied to the Sloan Digital Sky Survey
by \citet{Sousbie2008b}, providing by means of the length of the
skeleton, a good distinguishing tool for the analysis of the filamentary
structures. The formalism has recently been further developed and
applied to study the evolution of filamentary structure in simulations
\citep{Sousbie2009}.

Another approach is that of \citet{MMF07} (see also \citet{Ara2007c}).
They use the  Delaunay Triangulation Field Estimator (DFTE) to reconstruct
the density field for the galaxy distribution, and apply the
Multiscale Morphology Filter (MMF) to identify various structures, as
for instance clusters, walls, filaments and voids \citep{Ara2007b}.  As a further
development, this group has used the watershed algorithm to describe
the global properties of the density field \citep{Watershed08}.

A new direction is to use the second-order properties (the Hessian
matrix) of the density field \citep{Bond2009} or the deformation
tensor \citep{Klypin2008}. As is shown in these papers, this allows them
to trace and classify different features of the fields.

Our approach does not introduce the density estimation step; we
consider the galaxy distribution
as a marked point process.  In an earlier
paper~\citep{StoiMartMateSaar05}, we proposed to use an automated
method to trace filaments for realisations of marked point processes,
which has been shown to work well for the detection of road networks in
remote sensing
situations~\citep{LacoDescZeru05,StoiDescLiesZeru02,StoiDescZeru04}. This
method is based on the Candy model, a marked point process where
segments serve as marks. The Candy model can be applied to 2-D
filaments, and we tested it on simulated galaxy distributions. The
filaments we found delineated well the filaments detected by eye.

Based on our previous experience with the Candy model, we generalised
the approach for three dimensions. As the interactions between the
structure elements are more complex in three dimensions, we had to
define a more complex model, the Bisous model
\citep{StoiGregMate05}. This model gives a general framework for the
construction of complex patterns made of simple interacting
objects. In our case, it can be seen as a generalisation of the Candy
model. We applied the Bisous model to trace and describe the filaments
in the 2dFGRS \citep{StoiMartSaar07} and demonstrated that it works
well.

In the paper cited above we chose the observational samples from the
main magnitude-limited 2dFGRS catalogue, selecting the spatial regions
to have approximately constant spatial densities. But a strict application
of the Bisous process demands a truly constant spatial
density (intensity).  In this paper, we will apply the Bisous process
to compare the observational data with mock catalogues, specially
built to represent the 2dFGRS survey. To obtain strict
statistical test results, we use here volume-limited subsamples of the
2dFGRS and of the mock catalogues. We trace the filamentary network
in all our catalogues and compare its properties.

\section{Mathematical tools}
In this section we describe the main tools we use to study the
large-scale filaments. The key idea is to see this filamentary
structure as a realisation of a marked point process. Under this
hypothesis, the cosmic web can be considered as a random configuration
of segments or thin cylinders that interact, forming a network of
filaments. Hence, the morphological and quantitative characteristics
of these complex geometrical objects can be obtained by following a
straightforward procedure: constructing a model, sampling the
probability density describing the model, and, finally, applying the
methods of statistical inference.

We have given a more detailed description of these methods in a previous paper
\citep{StoiMartSaar07}. 

\subsection{Marked point processes}
A popular model for the spatial distribution of galaxies is a point
process on $K$ (a compact subset of $\RR^3$, the cosmologist's sample
volume), a random configuration of points $\kk=\{k_1,\ldots,k_n\}$,
lying in $K$. Let $\nu(K)$ be the volume of $K$.

We may associate characteristics or marks to the points. For instance,
to each point in a configuration $\kk$, shape parameters describing
simple geometrical objects may be attached. Let $(M,\cM,\nu_{M})$ be
the probability measure space defining these marks. A marked or object
point process on $K \times M$ is the random configuration $\yy
=\{(k_{1},m_{1}),(k_{2},m_{2}),\ldots,(k_{n},m_{n})\}$, with $y_i =
(k_i,m_i) \in K \times M$ for all $i=1,\ldots,n$ in a way that the
locations process is a point process on $K$. For our purposes, the
point process is considered finite and simple, {\it i.e.} the number
of points in a configuration is finite and $k_i \neq k_j$, for any
$i,j$ so that $1 \leq i,j \leq n$.

In case of the simplest marked point process, the objects do not
interact. The Poisson object point process is the most appropriate
choice for such a situation. This process chooses a number of objects
according to a Poisson law of the intensity parameter $\nu(K)$, gives
a random independent location to each object uniformly in $K$ and a
random shape or mark chosen independently according to $\nu_{M}$. The
Poisson object point process has the great advantage that it can be
described by analytical formulae. Still, it is too simple whenever
the interactions of objects are to be taken into account.

The solution to the latter problem is to specify a probability density
$p(\yy)$ that takes into account interactions between the
objects. This probability density is specified with respect to the
reference measure given by the Poisson object point process. There is
a lot of freedom in building such densities, provided that they are
integrable with respect to the reference measure and are locally
stable. This second condition requires that there exists $\Lambda >0$
so that $p(\yy \cup \{(k,m)\})/p(\yy) \leq \Lambda$ for any $(k,m)
\in K \times M$. Local stability implies integrability. It is also an
important condition, guaranteeing that the simulation algorithms for
sampling such models have good convergence properties.

For further reading and a comprehensive mathematical presentation of
object point processes, we recommend the monographs by \citet{Lies00},
\citet{MollWaag03}, \citet{StoyKendMeck95}, and \citet{Stoy08}.

\subsection{Bisous model}
In this section, we shall describe the probability density of the
Bisous model for the network of cosmic filaments. The Bisous model is
a marked point process that was designed to generate and analyse
random spatial patterns \citep{StoiGregMate05,StoiMartSaar07}.

Random spatial patterns are complex geometrical structures composed of
rather simple objects that interact. We can describe our problem as
follows: in a region $K$ of a finite volume, we observe a finite
number of galaxies $\dd=\{d_1,d_2,\ldots,d_r\}$. The positions of
these galaxies form a complicated filamentary network. Modelling it by
a network of thin cylinders that can get connected and aligned in a
certain way, a marked point process -- the Bisous model -- can be
built in order to describe it.

A random cylinder is an object characterised by its centre $k$ and its
mark giving the shape parameters. The shape parameters of a cylinder
are the radius $r$, the height $h$ and the orientation vector
$\omega$. We consider the radius and height parameters as fixed,
whereas the orientation vector parameters $\omega=\phi(\eta,\tau)$ are
uniformly distributed on $M=[0,2\pi) \times[0,1]$ so that
\begin{equation} 
\omega=(\sqrt{1-\tau^2}\cos(\eta),\sqrt{1-\tau^2}\sin(\eta),\tau). 
\end{equation}

For our purposes, throughout this paper the shape of a cylinder is
denoted by $s(y)=s(k,r,h,\omega)$, which is a compact subset of $\RR^{3}$ of
a finite volume $\nu(s(y))$. The shape of a random cylinder
configuration $\yy$ is defined by the random set $Z(\yy) =
\cup_{y \in \yy} s(y)$.
 
\begin{figure}
\centering
\resizebox{.8\hsize}{!}{\includegraphics*{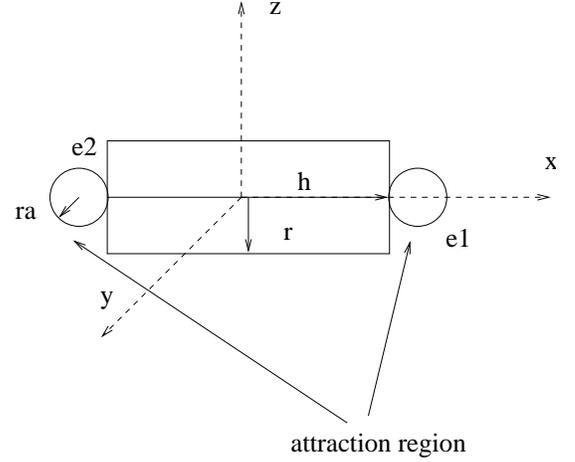}}
\caption{A thin cylinder that generates the filamentary network.}
\label{cylinder}
\end{figure}

A cylinder $(k,\omega)$ has $q=2$ extremity rigid points. We centre
around each of these points a sphere of the radius $r_a$. These two
spheres form an attraction region that plays an important role in
defining connectivity and alignment rules for cylinders. We illustrate
the basic cylinder in Fig.~\ref{cylinder}, where it is centred at the
coordinate origin and its symmetry axis is parallel to $Ox$. The
coordinates of the extremity points are
\begin{equation} 
e_{u} = ((-1)^{u+1}(\frac{h}{2}+r_a),0,0), \quad u \in \{1,2\} 
\end{equation} 
and the orientation vector is $\omega=(1,0,0)$.

The probability density for a marked point process based on random
cylinders can be written using the Gibbs modelling framework:
\begin{equation}
p(\yy | \theta) = \frac{\exp\left[-U(\yy|\theta)\right]}{\alpha}
\label{gibbs_density} 
\end{equation}
where $\alpha$ is the normalising constant, $\theta$ is the vector of
the model parameters and $U(\yy|\theta)$ is the energy function of the
system.

Modelling the filamentary network induced by the galaxy positions
needs two assumptions. The first assumption is that locally, galaxies
may be grouped together inside a rather small thin cylinder. The
second assumption is that such small cylinders may combine to extend a
filament if neighbouring cylinders are aligned in similar directions.

Following these two ideas the energy function given
by~(\ref{gibbs_density}) can be specified as: 
\begin{equation}
U(\yy|\theta) = U_{\dd}(\yy|\theta) + U_{i}(\yy|\theta)
\label{gibbs_energy}
\end{equation}
where $U_{\dd}(\yy|\theta)$ is the data energy and $U_{i}(\yy|\theta)$
is the interaction energy, associated to the first and second
assumptions above, respectively. In fact, it is perfectly reasonable
to think that the data energy is the reason that the cylinders in the galaxy
field are positioned just so, and that the interaction energy is the main 
factor which causes the cylinders to form filamentary patterns.

\subsection{Data energy}
The data energy of a configuration of cylinders $\yy$ is defined as
the sum of the energy contributions corresponding to each cylinder:
\begin{equation}
U_{\dd}(\yy|\theta) = - \sum_{y \in \yy} v(y)
\label{data_energy}
\end{equation}
where $v(\cdot)$ is the potential function associated to a cylinder
that depends on $\dd$ and the model parameters.

\begin{figure}
\centering
\resizebox{.7\hsize}{!}{\includegraphics{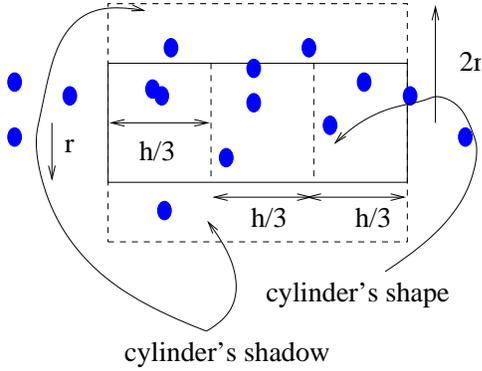}}
\caption{Two-dimensional projection of a thin cylinder with its
shadow within a pattern of galaxies.}
\label{data_cylinder}
\end{figure}

The cylinder potential is built taking into account local criteria
such as the density, spread and number of galaxies. To formulate these
criteria, an extra cylinder is attached to each cylinder $y$, with
exactly the same parameters as $y$, except for the radius which equals
$2r$. Let $\tilde{s}(y)$ be the shadow of $s(y)$ obtained by the
subtraction of the initial cylinder from the extra cylinder, as shown
in Fig.~\ref{data_cylinder}. Then, each cylinder $y$ is divided in
three equal volumes along its main symmetry axis, and we denote by
$s_{1}(y)$, $s_{2}(y)$ and $s_{3}(y)$ their corresponding shapes.

The local density condition verifies that the density of galaxies
inside $s(y)$ is higher than the density of galaxies in
$\tilde{s}(y)$, and it can be expressed as follows:
\begin{equation} 
n(\dd \cap s(y))/\nu(s(y)) > n(\dd \cap \tilde{s}(y))/\nu(\tilde{s}(y)),
\end{equation} 
where $n(\dd \cap s(y))$ and $n(\dd \cap \tilde{s}(y))$ are the 
numbers of galaxies covered by the cylinder and its shadow, 
and $\nu(s(y))$ and $\nu(\tilde{s}(y))$ are the volumes of the cylinder 
and its shadow, respectively. 

The even location of the galaxies 
along the cylinder main axis is ensured by the spread condition,
which is formulated as
\begin{equation} 
\prod_{i=1}^{3}n(\dd \cap s_{i}(y)) > 0,
\end{equation}
where $n(\dd \cap s_{i}(y))$ is the number of galaxies belonging to 
$s_{i}(y)$. 

If both these conditions are fulfilled, then $v(y)$ is given by the
difference between the number of galaxies contained in the cylinder and
the number of galaxies contained in its shadow:
\begin{equation} 
v(y)(\dd \cap s(y)) - n(\dd \cap \tilde{s}(y)).
\end{equation}
Whenever any of the
previous conditions is violated, a positive value $v_{\max}$ is
assigned to the potential of a cylinder.

A segment which does not fulfill the required conditions can still be
integrated into the network by the parameter $v_{\max}$. 
This should result in more complete networks and better mixing
properties to the method.

We note that we have chosen cylinders as the objects here in order to
trace filaments in the galaxy distribution. Such objects are tools at
our disposal and any object can be chosen; as an example,
\citet{StoiGregMate05} have built systems of flat elements (walls) and
of regular polytopes (galaxy clusters), based on the Bisous process.
 
\subsection{Interaction energy}

The interaction energy takes into account the interactions between
cylinders. It is the model component ensuring that the cylinders form
a filamentary network, and it is given by
\begin{equation} 
U_{i}(\yy|\theta) = - n_{\kappa}(\yy) \log \gamma_{\kappa} - \sum_{s=0}^{2}n_{s}(\yy)\log\gamma_{s}, 
\label{interaction_energy} 
\end{equation}  
where $n_{\kappa}$ is the number of repulsive cylinder pairs and $n_s$
is the number of cylinders connected to the network through $s$
extremity points. The variables $\log\gamma_{\kappa}$ and
$\log\gamma_s$ are the potentials associated to these configurations,
respectively.

\begin{figure}
\centering
\resizebox{.9\hsize}{!}{\includegraphics*{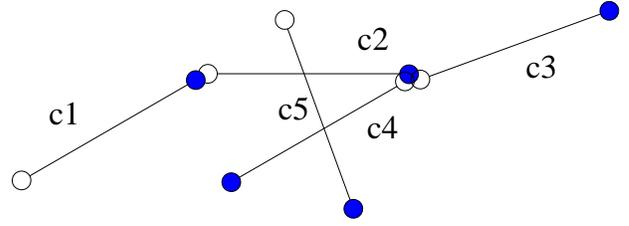}}
\caption{Two-dimensional representation of interacting cylinders.} 
\label{config_cylinder}
\end{figure}

We define the interactions that allow the configuration of cylinders
to trace the filamentary network below.  To illustrate these
definitions, we show an example configuration of cylinders (in two
dimensions) in Fig.~\ref{config_cylinder}.

Two cylinders are considered repulsive, if they are rejecting each
other and if they are not orthogonal. We declare that two cylinders
$y_{1}=(k_1,\omega_1)$ and $y_{2}=(k_2,\omega_2)$ reject each other if
their centres are closer than the cylinder height, $d(k_{1},k_{2}) <
h$. Two cylinders are considered to be orthogonal if $|\omega_{1}
\cdot \omega_{2}| \leq \tau_{\perp}$, where $\cdot$ is the scalar
product of the two orientation vectors and $\tau_{\perp} \in (0,1)$ is
a predefined parameter. So, we allow a certain range of mutual angles
between cylinders that we consider orthogonal.

Two cylinders are connected if they attract each other, do not
reject each other and are well aligned. Two cylinders attract
each other if only one extremity point of the first cylinder is
contained in the attraction region of the other cylinder. The
cylinders are ``magnetised'' in the sense that they cannot attract
each other through extremity points having the same index. Two
cylinders are well aligned if $\omega_{1} \cdot \omega_{2} \geq 1 -
\tau_{\parallel}$, where $\tau_{\parallel} \in (0,1)$ is a predefined
parameter.

Take now a look at Fig.~\ref{config_cylinder}.  According to the
previous definitions, we observe that the cylinders $c1$, $c_2$ and
$c_3$ are connected. The cylinders $c_1$ and $c_3$ are connected to
the network through one extremity point, while $c_2$ is connected to
the network through both extremity points. The cylinders $c_4$ and
$c_5$ are not connected to anything -- $c_4$ is not well aligned with
$c_2$, the angle between their directions is too large, and $c_5$ is
not attracted to any other cylinder. It is important to notice that
the cylinders $c_3$ and $c_4$ are not interacting -- they are wrongly
'polarised', their overlapping extremity points have the same
index. The cylinder $c_5$ is rejecting the cylinders $c_2$ and $c_4$
(the centres of these cylinders are close), but as it is rather
orthogonal both to $c2$ and $c_4$, it is not repulsing them. The
cylinders $c_2$ and $c_4$ reject each other and are not orthogonal, so
they form a repulsive pair.

Altogether, the configuration at Fig.~\ref{config_cylinder} adds to
the interaction energy contributions from three connected cylinders
(one doubly-connected, $c_2$, and two single-connected, $c_1$ and
$c_3$), and from one repulsive cylinder pair ($c_2$--$c_4$).

The complete model~(\ref{gibbs_density}) that includes the definitions
of the data energy and of the interaction energy given
by~(\ref{data_energy}) and~(\ref{interaction_energy}) is well defined
for parameters as $v_{\max} < 0$, $\gamma_0,\gamma_1,\gamma_2 >
0$ and $\gamma_\kappa \in [0,1]$. The definitions of the interactions
and the parameter ranges chosen ensure that the complete model is
locally stable and Markov in the sense of
Ripley-Kelly~\citep{StoiGregMate05}. For cosmologists it means that
we can safely use this model without expecting any dangers (numerical,
convergence, etc.).

\subsection{Simulation}
Several Monte Carlo techniques are available to simulate marked point
processes: spatial birth-and-death processes, Metropolis-Hastings
algorithms, reversible jump dynamics or more recent exact simulation
techniques~\citep{GeyeMoll94,Geye99,Gree95,KendMoll00,Lies00,LiesStoi06,Pres77}.

In this paper, we need to sample from the joint probability density
law $p(\yy,\theta)$.  This is done by using an iterative Monte Carlo
algorithm. An iteration of the algorithm consists of two steps. First,
a value for the parameter $\theta$ is chosen with respect to
$p(\theta)$. Then, conditionally on $\theta$, a cylinder pattern is
sampled from $p(\yy|\theta)$ using a Metropolis-Hastings
algorithm~\citep{GeyeMoll94,Geye99}.

The Metropolis-Hastings algorithm for sampling the conditional law
$p(\yy|\theta)$ has a transition kernel based on three types of
moves. The first move is called {\it birth} and proposes to add a
new cylinder to the present configuration. This new cylinder can be
added uniformly in $K$ or can be randomly connected with the rest of
the network. This mechanism helps to build a connected network. The
second move is called {\it death}, and proposes to eliminate a
randomly chosen cylinder. The role of this second move is to ensure
the detailed balance of the simulated Markov chain and its convergence
towards the equilibrium distribution. A third move can be added to improve 
the mixing properties of the sampling algorithm . This
third move is called {\it change}; it randomly chooses a cylinder in
the configuration and proposes to ``slightly '' change its parameters
using simple probability distributions. For specific details
concerning the implementation of this dynamics we
recommend~\citet{LiesStoi03} and \citet{StoiGregMate05}.

Whenever the maximisation of the joint law $p(\yy,\theta)$ is needed,
the previously described sampling mechanism can be integrated into a
simulated annealing algorithm. The simulated annealing algorithm is
built by sampling from $p(\yy,\theta)^{1/T}$, while $T$ goes slowly to
zero. \citet{StoiGregMate05} proved the convergence of such simulated
annealing for simulating marked point processes, when a logarithmic
cooling schedule is used. According to this result, the temperature is
lowered as
\begin{equation}
T_{n} = \frac{T_{0}}{\log n+1};
\label{cooling_schedule}
\end{equation}
we use $T_{0}=10$ for the initial temperature.

\subsection{Statistical inference}
One straightforward application of the simulation dynamics is the
estimation of the filamentary structure in a field of galaxies
together with the parameter estimates. These estimates are given by~:
\begin{eqnarray} 
(\widehat{\yy},\widehat{\theta}) & = & \arg\max_{\Omega \times \Psi} p(\yy,\theta)= \arg\max_{\Omega \times \Psi} p(\yy|\theta)p(\theta) \nonumber \\
& = &\arg\min_{\Omega \times \Psi}\left\{ \frac{U_{\dd}(\yy|\theta)+U_{i}(\yy|\theta)}{\alpha(\theta)} + \frac{U_{p}(\theta)}{\alpha_{p}(\theta)}\right\}, 
\label{estimator_pattern} 
\end{eqnarray} 
where $\alpha(\theta)$ is the normalising constant,
$p(\theta)=\exp[-U_{p}(\theta)]/\alpha_{p}(\theta)$ is the prior law
for the model parameters and $\Psi$ is the model parameter's space.
 
However, the solution we obtain is not unique. In practice, 
the shape of the prior law $p(\theta)$ may influence the solution,
making the result to look more random compared with a result obtained
for fixed values of parameters. Therefore, it is reasonable to wonder
how precise the estimate is, that is if an element of the pattern
really belongs to the pattern, or if its presence is due to random
effects~\citep{StoiGayKret07,StoiMartSaar07})
 
For compact subregions $\cR \subseteq K$, 
we can compute or give Monte Carlo approximations
for average quantities such as
\begin{equation} 
\mathrm{E}_{(\YY,\Theta)}\left[f(\cR, Z(\YY))\right],
\label{expectation} 
\end{equation}
\noindent
where $\mathrm{E}$ denotes the expectation value over the data and
model parameter space, and $f(\cR,\cdot)$ is a real measurable
function with respect to the $\sigma$-algebra associated to the
configuration space of the marked point process.

If $f(\cR, Z(\YY)) = \1\{\cR \subseteq Z(\YY)\}$ (where $\1$ is the
indicator function), then the expression~(\ref{expectation})
represents the probability of how often the considered model includes
or visits the region $\cR$. Furthermore, if $K$ is partitioned into a
finite collection of small disjoint cells
$\{\cR_1,\cR_2,\ldots,\cR_q\}$, then a visit probability map can be
obtained. This map is given by the partition together with the value
$P_{i} = \mathrm{E} \left[ \1\{\cR_i \subseteq Z(\YY)\}\right]$
associated to each cell. The map is defined by the model and by the
parameters of the simulation algorithm; its resolution is given by the
cell partition.

The sufficient statistics of the model~(\ref{interaction_energy}) --
the interaction parameters $n_{\kappa}$ and $n_{s}, s=(0,1,2)$ --
describe the size of the filamentary network and quantify the
morphological properties of the network. Therefore, they are suitable
as a general characterisation of the filamentarity of a galaxy
catalogue. This renders the comparison of the networks of different regions and/or
different catalogues perfectly possible. Here, we use the
sufficient statistics to characterise the real data and the mock
catalogues.

The visit maps show the location and configuration of the filament
network.  Still, the detection of filaments and this verification test
depend on the selected model. It is reasonable to ask if these results
are obtained because the data exhibits a filamentary structure or just
because of the way the model parameters are selected.

The sufficient statistics can be used to build a statistical test in
order to answer the previous question. For a given data catalogue,
samples of the model are obtained, so the means of the sufficient
statistics can be then computed. The same operation, using exactly the
same model parameters, can be repeated whenever an artificial point
field -- or a synthetic data catalogue -- is used. If the artificial
field is the realisation of a binomial point process having the same
number of points as the number of galaxies in the original data set,
the sufficient statistics are expected to have very low values --
there is no global structure in a such binomial field. If the values
of the sufficient statistics for these binomial fields were large,
this would mean that the filamentary structure is due to the
parameters, not to the data. Comparing the values obtained for the
original data sets with Monte Carlo envelopes found for artificial
point fields, we can compute Monte Carlo $p$-values to test the
hypothesis of the existence of the filamentary structure in the
original data catalogue~\citep{StoiGayKret07,StoiMartSaar07}.

\section{Data}

We apply our algorithms to a real data catalogue and compare the
results with those obtained for 22 mock catalogues, specially
generated to simulate all main features of the real data.

\begin{figure*}
\centering
\resizebox{.96\hsize}{!}{\includegraphics*{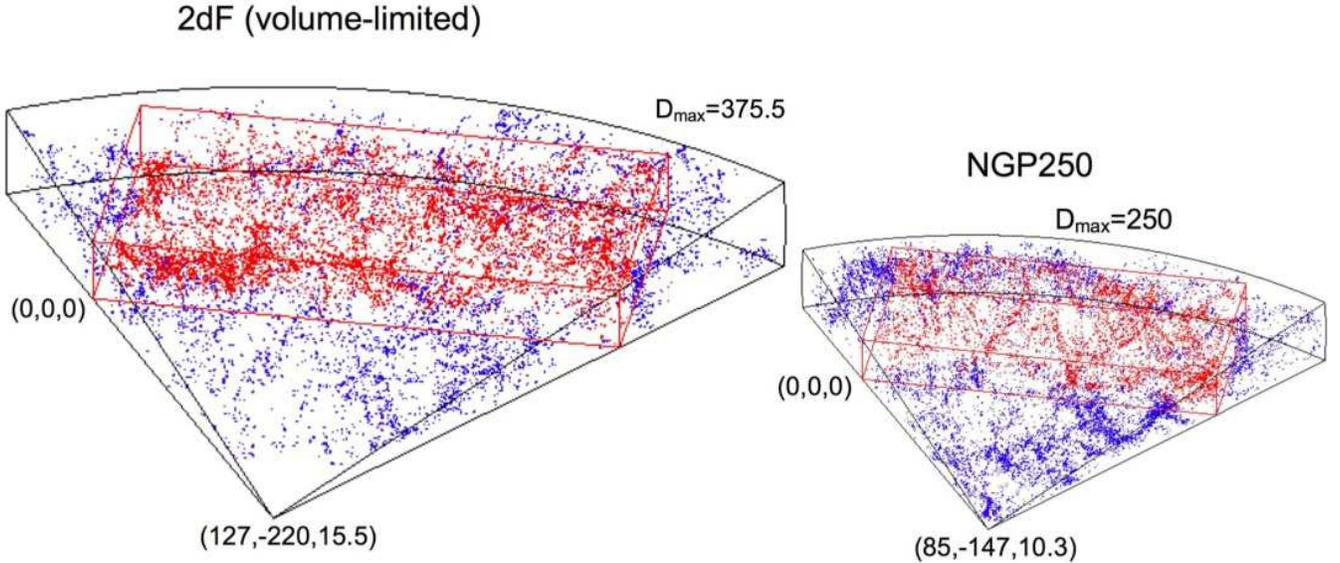}}
\caption{The geometry and coordinates of the data bricks. Right panel: all 2dF galaxies inside a large contiguous area
of the northern wedge are shown in blue (up to a depth of 250 $h^{-1}$Mpc), galaxies that belong to the NGP250 sample are depicted in 
red. The left panel corresponds to the volume-limited sample. Both diagrams are shown to scale; 
there is very little overlap (only $15.6h^{-1}$Mpc in depth)
between the NGP250 and the 2dF volume-limited bricks. The coordinates are
in units of 1$h^{-1}$Mpc.}
\label{fig:data}
\end{figure*}

\subsection{Observational data}

At the moment there are two large galaxy redshift (spatial position)
catalogues that are natural candidates for a filament search.  When the
work reported here was carried out a few years ago, the best available
redshift catalogue to study the morphology of the galaxy distribution
was the 2 degree Field Galaxy Redshift Survey
\citep[2dFGRS,][]{2dFGRS}; the much larger Sloan Digital Sky Survey
(SDSS) \citep[see the description of its final status in][]{SDSS7} was
yet in its first releases. Also, only the 2dFGRS had a collection at that time
of mock catalogues that were specially generated to mimic
the observed data. So this study is based on the 2dFGRS; we shall
certainly apply our algorithms to the SDSS in the future, too.

The 2dFGRS covers two separate regions in the sky, the NGP (North
Galactic Cap) strip, and the SGP (South Galactic Cap) strip, with a
total area of about 1500 square degrees. The nominal
(extinction-corrected) magnitude limit of the 2dFGRS catalogue is
$b_j=19.45$; reliable redshifts exist for 221,414 galaxies.  The
effective depth for the catalogue is about $z=0.2$ or a comoving
distance of $D=572\,h^{-1}\,$Mpc for the standard cosmological model
with $\Omega_{\rm matter}=0.3$ and $\Omega_{\Lambda}=0.7$\footnote{Here 
and below $h$ is the
dimensionless Hubble constant, $H=h\cdot100$ km s$^{-1}$ Mpc$^{-1}$.}.

The 2dFGRS catalogue is a flux-limited catalogue and therefore the
density of galaxies decreases with distance. For a statistical analysis
of such surveys, a weighting scheme that compensates for the missing
galaxies at large distances has to be used.  However, such a
weighting is suitable only for specific statistical problems, as
e.g. the calculation of correlation functions. When studying the
local structure, such a weighting cannot be used; it would only
amplify the shot noise.
 
We can eliminate weighting by using volume-limited samples.  The 2dF
team has generated these for scaling studies \citep[see,
  e.g.,][]{croton1}; they kindly sent these samples to us. The
volume-limited samples are selected in one-magnitude intervals; we
chose as our sample the one with the largest number of galaxies for the
absolute magnitude interval M$_b\in[-19.0,-20.0]$. The total number of
galaxies in this sample is 44,713.

The NGP250 sample is good for detecting filaments, as shown in our
previous paper \citep{StoiMartSaar07} . But this sample is 
magnitude-limited (not volume-limited), therefore the number of galaxies 
decreases with depth, because only galaxies with an apparent magnitude
exceeding the survey cutoff are detected.
Since we can perform statistical tests only when our base point process is
a Poisson process, implying approximately constant mean density with depth, we
have to use volume-limited samples in our study.
Moreover, the mocks that have been built for the 2dFGRS are already volume-limited, 
and cannot be combined into a magnitude-limited sample because of their 
different depths. Thus, if we want to compare the observed filaments with those in the
mock samples, we are forced to use volume-limited samples.
 
The borders of the two volumes covered by the sample are rather
complex. As our algorithm is recent, we do not yet have the estimates of
the border effects, and we cannot correct for these. So we limited our
analysis to the simplest volumes -- bricks.  As the southern half of
the galaxy sample has a convex geometry (it is limited by two conical
sections of different opening angles), the bricks which are possible to
cut from there have small volumes. Thus we used only the northern data
which have a geometry of a slice, and chose the brick of a maximum
volume that could be cut from the slice. We will compare the
results obtained for this sample (2dF) below  with those obtained for a smaller
sample in a previous paper (NGP250); the geometry and galaxy content
of these two data sets is described in Table~\ref{tab:bricks}. 
We have shifted the origin of the coordinates to the near lower left
corner of the brick; the geometry of the bricks (both the 2dF and
NGP250 sample) is illustrated in Fig.~\ref{fig:data}. 

\begin{table}
\centering
\setlength{\tabcolsep}{4pt}
\begin{tabular}{rrrrrr} 
\hline\\ 
sample&$N_{\mbox{gal}}$&depth&width&height&$d$\\[3pt] 
\hline\\ 
2dF&8487&133.1&254.0&31.1&5.0\\
NGP250&7588&88.6&169.1&20.7&3.4\\[3pt] 
\hline\\ 
\end{tabular}
\caption{Galaxy content and geometry for the data bricks (sizes are in
$h^{-1}$Mpc). $N_{\mbox{gal}}$ is the number of galaxies in the sample, 
and $d$ is the mean
distance between galaxies in the sample.}
\label{tab:bricks} 
\end{table} 

\subsection{Mock catalogues}
We compare the observed filaments with those built for mock galaxy
catalogues which try to simulate the observations as closely as
possible. The construction of these catalogues is described in detail
by \citet{Norberg02}; we give a short summary here.  The 2dF mock
catalogues are based on the ``Hubble Volume'' simulation
\citep{HubbleVol}, an N-body simulation of a 3$h^{-1}$~Gpc cube of
$10^9$ mass points.  These mass points are considered as galaxy
candidates and are sampled according to a set of rules that include:
\begin{enumerate}
\item Biasing: the probability for a galaxy to be selected is calculated on the basis of the smoothed (with a $\sigma=2h^{-1}\mbox{Mpc}$ Gaussian filter) final density. This probability (biasing) is exponential \citep[rule 2 of][]{Cole98}, with parameters chosen to reproduce the observed power spectrum of galaxy clustering.
\item Local structure: the observer is placed in a location similar to our local cosmological neighbourhood.
\item A survey volume is selected, following the angular and distance selection factors of the real 2dFGRS.
\item Luminosity distribution: luminosities are assigned to galaxies according to the observed (Schechter) luminosity distribution; $k+e$-corrections are added.
\end{enumerate}
These ``ideal'' catalogues are then combined with observational errors to produce the final mock catalogues:
\begin{enumerate}
\item Galaxy redshifts are modified by adding random dynamical velocities.
\item Observational random errors are added to galaxy magnitudes.
\item Based on galaxy positions, survey incompleteness factors are calculated.
\end{enumerate}
These catalogues are as close to the observed catalogues as currently possible -- the spatial coverage, galaxy density, clustering, luminosities and observational errors are the same. So, we expect that the filamentary structure of the mock catalogues should be close to the ones we observe.

\begin{figure*}
\centering
\resizebox{0.45\textwidth}{!}{\includegraphics*{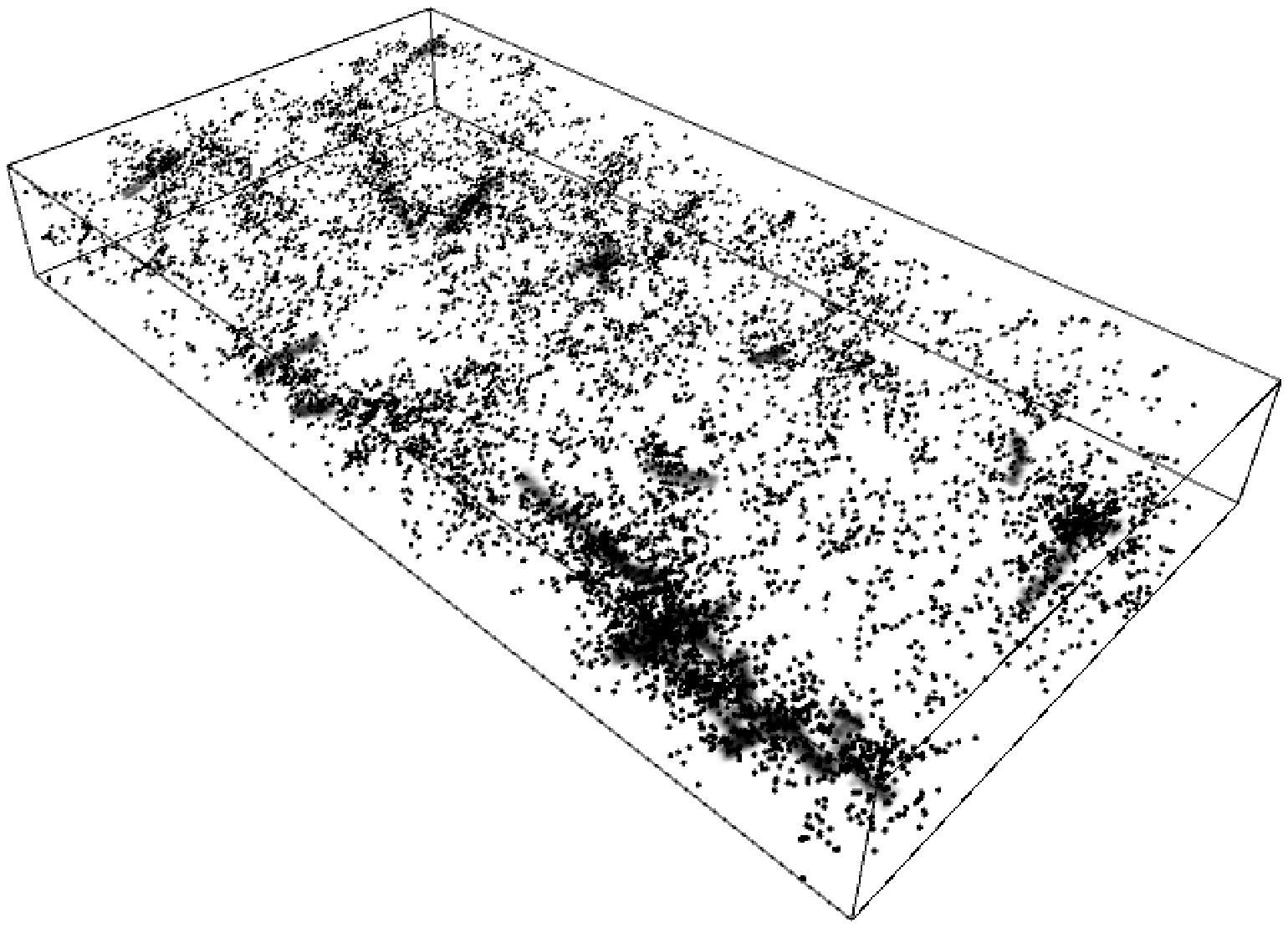}}
\resizebox{0.45\textwidth}{!}{\includegraphics*{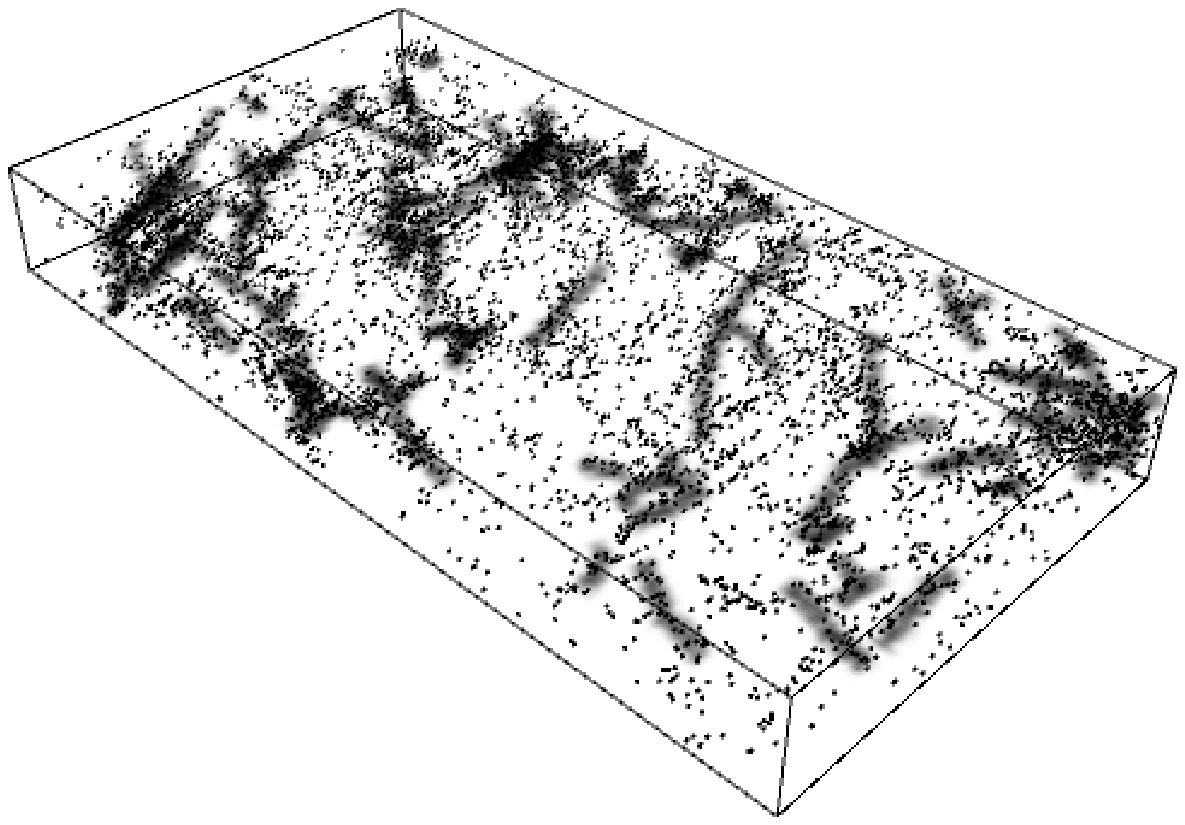}}\\
\caption{Filaments in the main data set 2dF (left panel) and in a
  smaller, but more dense data set N250 (right panel). The volumes are
  shown at the same scale.}
\label{fig:2df_ann}
\end{figure*}

\section{Filaments}

\subsection{Experimental setup}
As described above, we use the data sets drawn from the galaxy
distribution in the Northern subsample of the 2dFGRS survey and from
the 22 mock catalogues. For mock catalogues, we use the same absolute
magnitude range and cut the same bricks as for the 2dFGRS survey.

The sample region $K$ is the brick. In order to choose the values for
the dimensions of the cylinder we use the physical dimensions of the
galaxy filaments that have been observed in more detail
\citep{pimbblet04a}; we used the same values also in our previous
paper \citep{StoiMartSaar07}: a radius $r=0.5$ and a height $h=6.0$
(all sizes are given in $h^{-1}$Mpc). The radius of the cylinder is
close to the minimal one can choose, taking into account the data
resolution.  Its height is also close to the shortest possible, as our
shadow cylinder has to have a cylindrical geometry, too (the ratio of
its height to the diameter is presently 3:1). We choose the attraction
radius as $r_{a}=0.5$, giving the value 1.5 for the maximum distance
between the connected cylinders, and for the cosines of the maximum
curvature angles we choose $\tau_{\parallel}=\tau_{\perp}=0.15$. This
allows for a maximum of $\approx30^\circ$ between the direction angles
of connected cylinders and considers the cylinders to be orthogonal, if the
angle between their directions is larger than $\approx80^\circ$.

The model parameters $(r,h,r_{a})$ influence the detection results. If
they are too low, all network will be considered as made of
clusters, so no filaments will be detected. If they are too high, the
detected filaments will be too wide and/or too sparse, and precision
will be lost. Still, this makes the visit maps an interesting tool,
since, in a certain manner, they average the detection result. In this
work, the $(r,h,r_{a})$ parameters were fixed after a visual
inspection of the data and of different projections outlining the
filaments.

The marked point process-based methodology allows us to introduce these
parameters as marks or priors characterised by a probability density,
hence the detection of an optimal value for these parameters is then
possible. Knowledge based on astronomical observations could be used
to set the priors for such probability densities.

For detecting the scale-length of the cylinders or for obtaining
indications about its distribution, we may use visit maps to build
cell hypothesis tests to see which the most probable $h$ of the
cylinder passing through this cell could be. This may require also a refinement of
the data term of the model.
 
For the data energy, we limit the parameter domain by
$u_{\max}=[-25,20]$. For the interaction energy, we choose the
parameter domain as follows: $\log \gamma_{0} \in [-12.5,-7.5]$, $\log
\gamma_{1} \in [-5, 0] $ and $\log \gamma_{2} \in [0, 5]$. The hard
repulsion parameter is $\gamma_{k}=0$, so the configurations with
repulsing cylinders are forbidden. The domain of the connection
parameters was chosen in a way that $2$-connected cylinders are generally
encouraged, $1$-connected cylinders are penalised and $0$-connected
segments are strongly penalised. This choice encourages the cylinders
to group in filaments in those regions where the data energy is good
enough.  Still, we have no information about the relative strength of
those parameters.  Therefore, we have decided to use 
the uniform law over the parameter
domain for the prior parameter density $p(\theta)$.

\subsection{Observed filaments}
We ran the simulated annealing algorithm for 250,000 iterations;
samples were picked up every 250 steps. 
 
The cylinders obtained after running the simulated annealing outline
the filamentary network. But as simulated annealing requires
an infinite number of iterations till convergence, and also because of the
fact that an infinity of solutions is proposed (slightly changing the
orientation of cylinders gives us another solution that is as good as
the original one), we shall use visit maps to ``average" the shape of
the filaments.

Figure \ref{fig:2df_ann} shows the cells that have been visited by
our model with a frequency higher than 50\%, together with the galaxy
field. Filamentary structure is seen, but the filaments tend to be
short, and the network is not very well developed.  For comparison, we
show a similar map for the smaller volume (NGP250), where the galaxy
density is about three times higher. We see that the effectiveness of
the algorithm depends strongly on the galaxy density; too much of a
dilution destroys the filamentary structure.
 
As galaxy surveys have different spatial densities, this problem should be addressed.
The obvious way to do that is to rescale the basic cylinder. First, we can do full parameter estimation, with cylinder sizes included. Second, we can use an empirical approach, choosing a few nearby well-studied filaments, removing their fainter galaxies and finding the values for $h$ and $r$ that are needed to keep the filaments together. 

\begin{figure}
\centering
\resizebox{0.45\textwidth}{!}{\includegraphics*{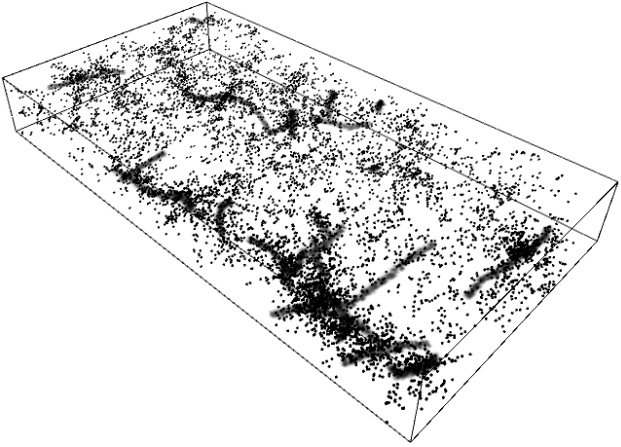}}
\caption{Filaments in the main data set 2dF, for the rescaled basic cylinder.} 
\label{fig:2dfL_ann}
\end{figure}

But this needs a separate study. We will use here a simple density-based rescaling -- as the density of the 2dF sample is three times lower than that of the smaller volume
we rescaled the cylinder
dimensions by $3^{1/3}=1.44$.  The filamentary network for this case
is shown in Fig.~\ref{fig:2dfL_ann}. This is better developed, but not
as well delineated as that for the smaller volume. 

This rescaling assumes that the dilution is Poissonian, and 
there is no luminosity-density relation. Both those assumptions are wrong. Our justification of the scaling used here is that this is the simplest scaling assumption,
and the dimensions of the rescaled cylinder ($r=0.72, h=8.6$) do not contradict observations. Also, the filamentary networks found with rescaled cylinders and
the visit maps seem to better trace the filaments seen by eye. We realise that the scaling problem is important, and will return to  it in the future.

As we work in the redshift space, the apparent galaxy distribution is distorted
by peculiar velocities in groups and clusters that produce so-called
'fingers-of-god', structures that are elongated along the line-of-sight.
These fingers may masquerade as filaments for our procedure. To estimate
their influence, we first found the cylinders
using the simulated annealing algorithm. The 
cylinders along the line-of-sight may
be caused by the finger-of-god effect. A simple
test was implemented, checking if the module of the scalar product
between the direction of the symmetry axis of the cylinder and the
direction of the line-of-sight ($|\cos\phi|$, where $\phi$ is the angle between these
directions) is close to $1$ (greater than
$0.95$). 

\begin{figure*}
\centering
\resizebox{.48\hsize}{!}{\includegraphics*{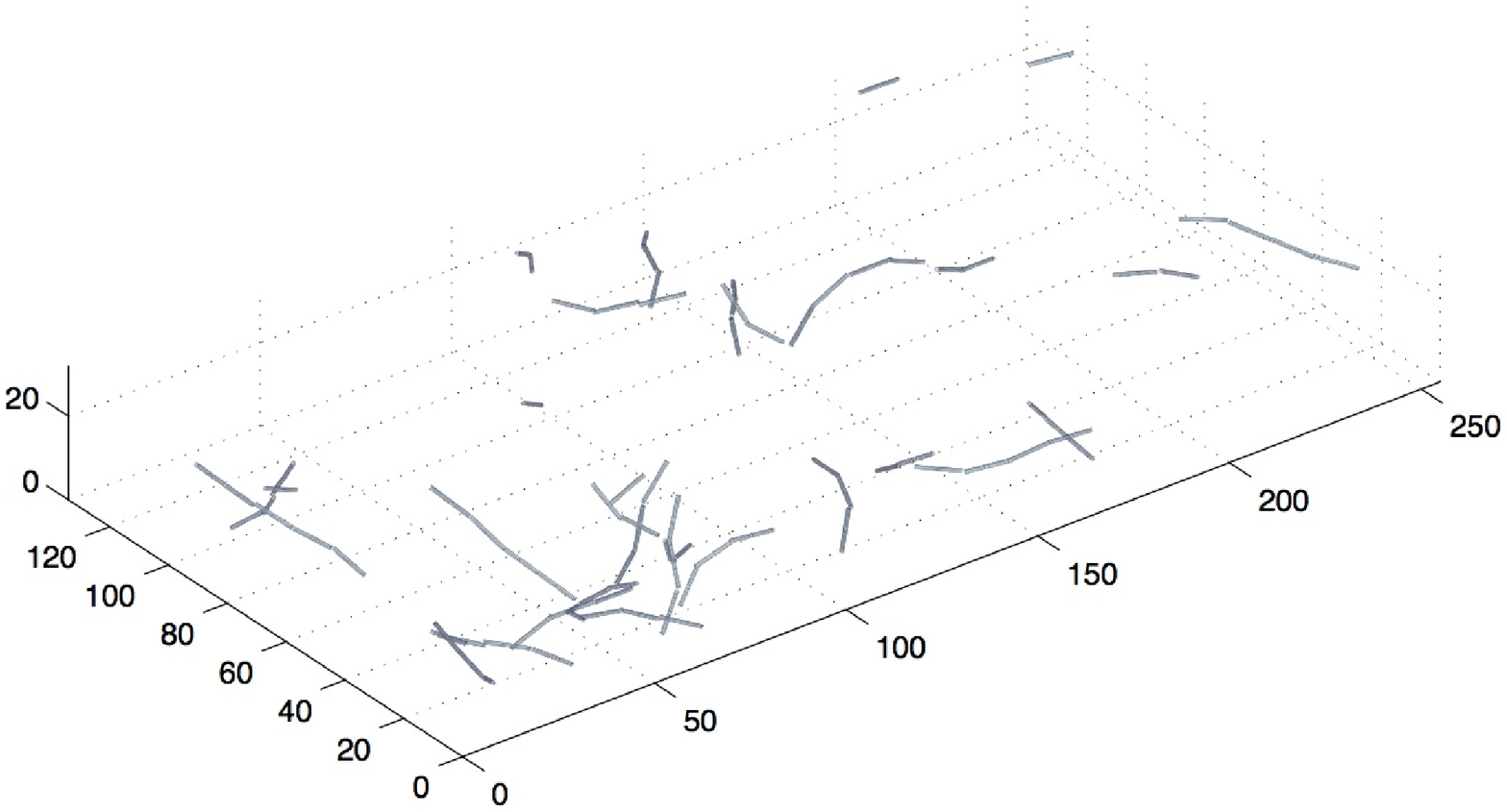}}
\resizebox{.48\hsize}{!}{\includegraphics*{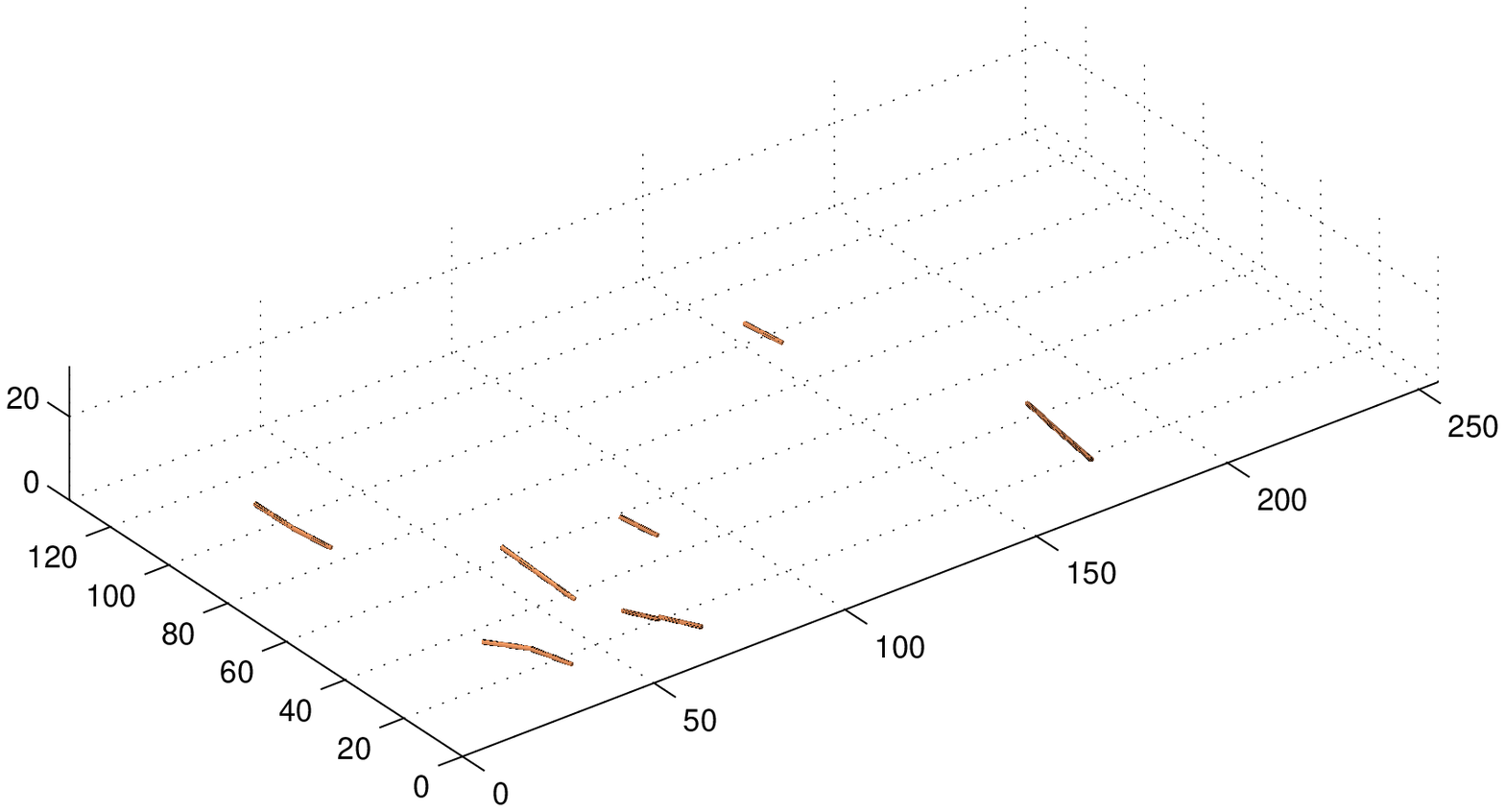}}\\
\caption{The full cylinder network (left panel) and line-of-sight cylinders
(right panel), for the 2dF data brick, with a rescaled cylinder. The coordinates are
in units of 1$h^{-1}$Mpc.}
\label{fig:fingers}
\end{figure*}

The results are shown in Table~\ref{tab:fingers} and in Fig.~\ref{fig:fingers}.
The Table shows the total number of cylinders $n_t$, the number of
line-of-sight cylinders $n_f$, and the expected number of cylinders $n_e$
(assuming an isotropic distribution of cylinders, $n_e=0.05n_t$).
Figure~\ref{fig:fingers} compares the network of all cylinders (left panel)
with the location of the line-of-sight cylinders (right panel). The figures for
all other catalogues listed in Table~\ref{tab:fingers} appear to be similar.

\begin{table} 
\centering
\begin{tabular}{l|ccc}
Data & $n_t$ & $n_f$ & $n_e$\\
\hline
MOCK8 (A) & 57 & 5 & 2.9\\
MOCK8  & 30 & 5 & 3.0\\
MOCK16 (A) & 107 & 24 & 5.4\\
MOCK16  & 82 & 18 & 4.1\\
2dF (A) & 86 & 13 & 4.3\\
2dF  & 65 & 14 & 3.3\\
NGP 250  & 191 & 21 & 9.6\\
\hline
\end{tabular}
\caption{Line-of-sight cylinders in the data (2dF and NGP250) and in two mocks, 8 and 16. 
The index (A) labels the rescaled case with a larger cylinder. The column $n_t$ shows
the total number of cylinders in the network, $n_f$ is the number of line-of-sight
cylinders, and $n_e$ is the expected number of line-of-sight cylinders, in case of
the isotropic cylinder orientation.
}
\label{tab:fingers} 
\end{table} 

Clearly, our method detects such fingers, although the number of extra cylinders
(fingers-of-god) is not large. There are at least two possibilities to exclude them.
The first is to use a group catalogue, where fingers are already compressed,
instead of a pure redshift space catalogue. Another possibility is to
modify our data term, checking for the cylinder orientation,
and to eliminate the fingers within the algorithm.
We will test both possibilities in future work.

A problem that has been addressed in most of the papers about galaxy
filaments is the typical filament length (or the length
distribution). As our algorithm allows branching of filaments
(cylinders that are approximately orthogonal), it is difficult to
separate filaments. We have tried to cut the visit maps into filaments,
but the filaments we find this way are too short. 
We may advance here applying morphological operations to
visit maps. Still, this kind of operation needs at
least a good mathematical understanding of the ``sum'' of all the
cells forming the visit maps.

Another possibility is to
use the cylinder configurations for selecting individual filaments.
These configurations can be thought of as a (somewhat random) filament
skeletons of visit maps. We have used them to find the distributions of the sufficient
statistics, and these configurations should be good enough to estimate
other statistics, as filament lengths. We will certainly try that in the future.

There are
problems where knowing the typical filament length is very important,
as in the search for missing baryons.  These are
thought to be hidden as warm intergalactic gas (WHIM, see, e.g.\citet{Viel05}).
In order to detect this gas, the best candidates are galaxy filaments that
lie approximately along the line-of-sight; knowing the typical length of a
filament we can predict if a detection would be possible.

Concerning the length of the entire filamentary network, the most
direct way of estimating it is to multiply the number of cylinders
with $h$. In this case, the distribution of the length of the network
is given by the distribution of the sum of the three sufficient
statistics of the model. The precision of the estimator is related to
the precision of $h$. Another possible estimator can be constructed
using $h+r_c$ instead of $h$. The same construction can be used even
if different cylinders are
used~\cite{LacoDescZeru05,StoiDescLiesZeru02,StoiDescZeru04}. As for
the sufficient statistics, the distribution of the length of the
network may be derived using Monte Carlo techniques.

We note that we can easily find the total volumes
of filaments, counting the cells on the visiting map. 
As an example, for the
cases considered here, the relative filament volumes are $1.8\cdot10^{-3}$
(2dF, smaller cylinder), $3.3\cdot10^{-3}$ (2dF, rescaled cylinder),
and $1.6\cdot10^{-2}$ (NGP250).

\subsection{Statistics}
\begin{figure*}
\centering
\resizebox{.33\hsize}{!}{\includegraphics*{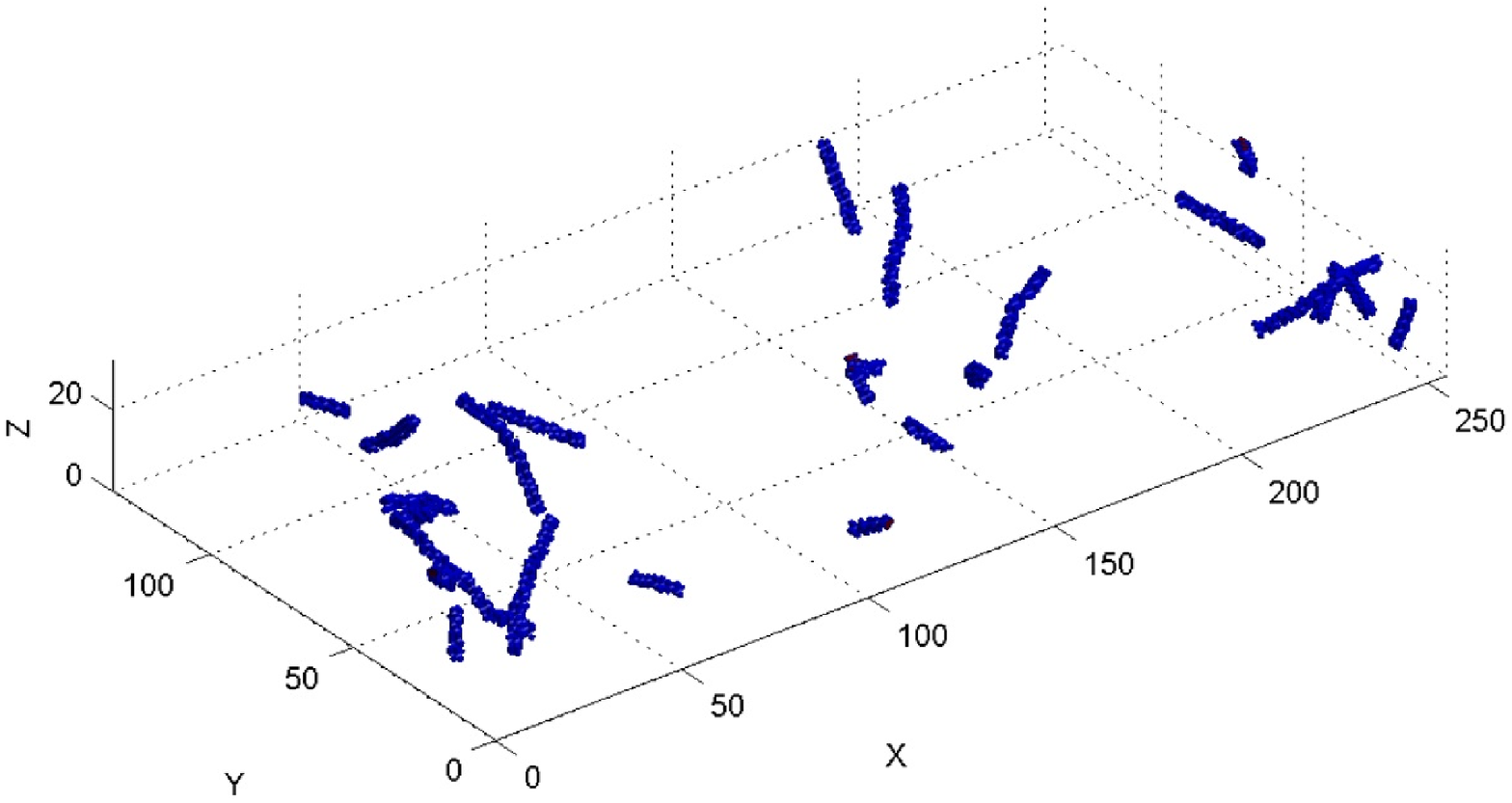}}
\resizebox{.33\hsize}{!}{\includegraphics*{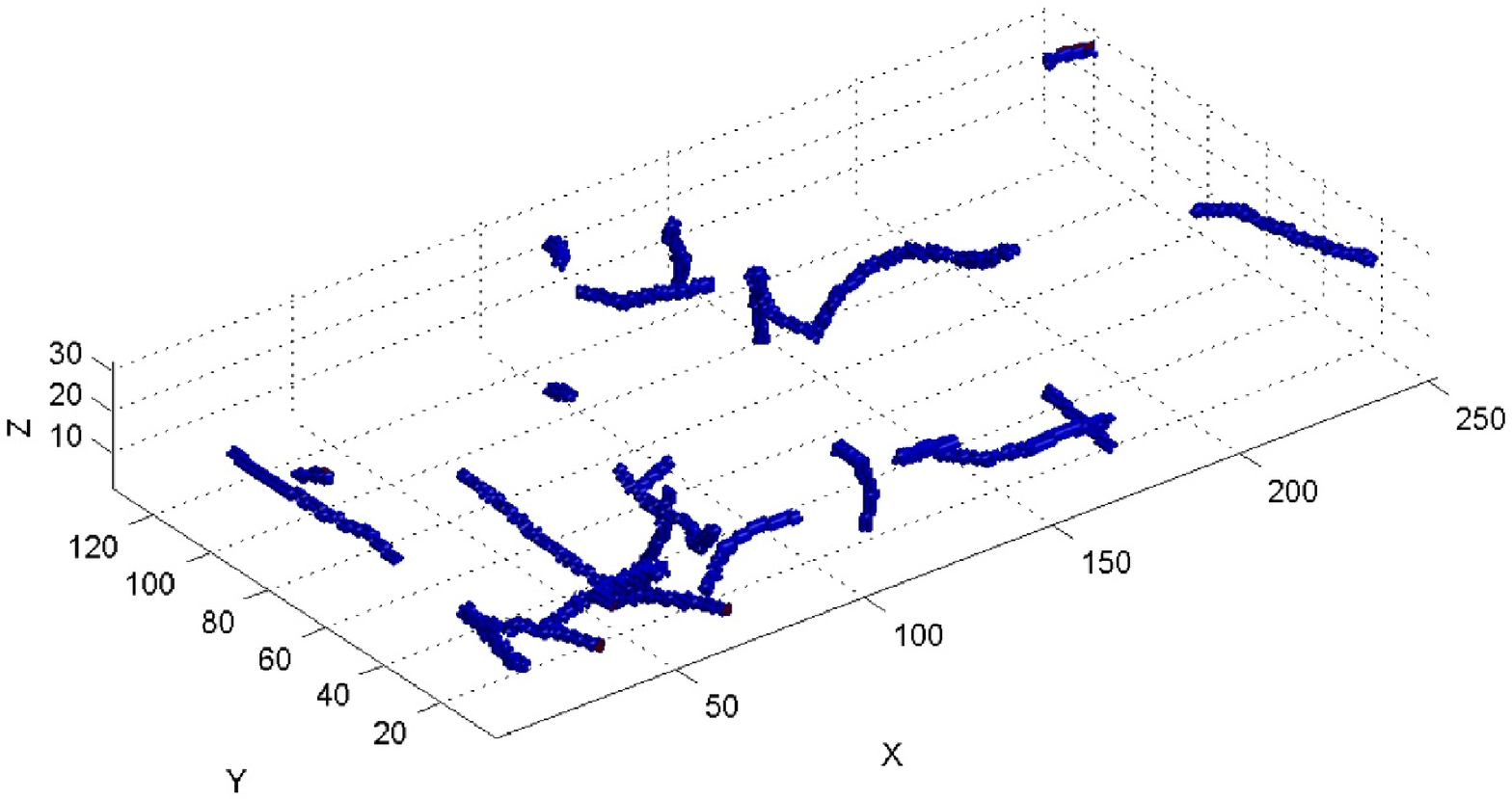}}
\resizebox{.33\hsize}{!}{\includegraphics*{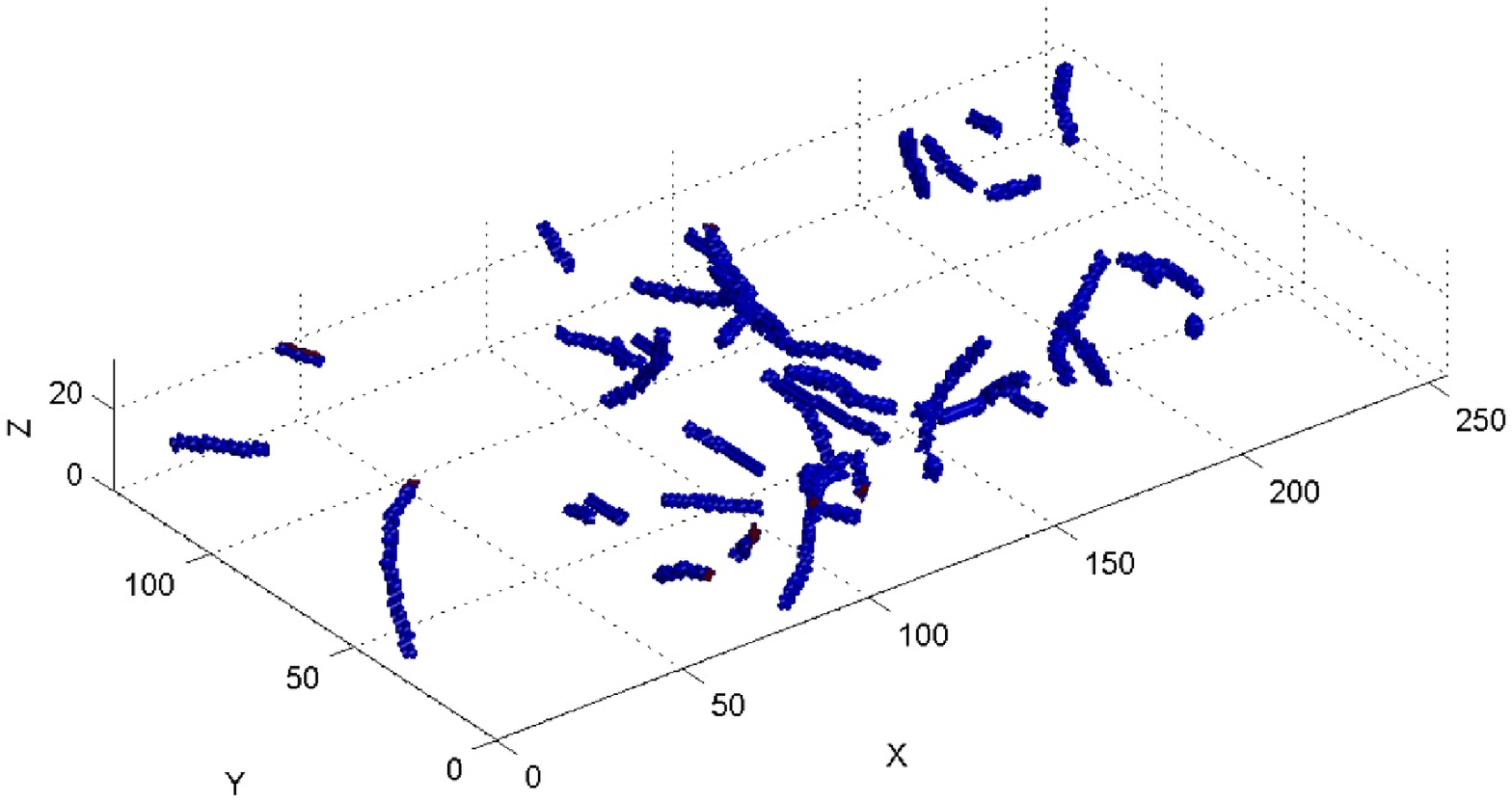}}\\
\caption{Visit maps for two mocks (mock8 -- left panel, mock16 -- right panel) and the real data (middle panel), for the rescaled cylinder. We show the cells for the 50\% threshold.}
\label{fig:visits}
\end{figure*}

As we explained before, in order to compare the filamentarity of the
observed data set (2dF) and the mocks, we had to run the
Metropolis-Hastings algorithm at a fixed temperature $T=1.0$ (sampling
from $p(\yy,\theta)$).  The algorithm was run for $250,000$ iterations,
and samples were picked up every $250$ iterations. The means of the
sufficient statistics of the model were computed using these
samples. As an example, we compare
the single-temperature visit map for the data with two extreme cases
for the mocks (8 and 16) in Fig.~\ref{fig:visits}.
The obtained results are shown in Table~\ref{tab:statdata}.

\begin{table}
\centering
 \begin{tabular}{|l|r|r|r|} 
\hline 
& \multicolumn{3}{c|}{\strut Sufficient statistics}\\ 
\cline{2-4} \raisebox{1ex}[0pt]{Data sets}&\strut  $\bar{n_2}$ & $\bar{n_0}$  
& $\bar{n_1}$ \\ \hline 
\strut 2dF & 1.94  & 5.30 & 11.66\\ 
MOCK 1 & 2.53  & 5.62 & 13.16\\ 
MOCK 2 & 0.48 & 6.20 & 7.52\\ 
MOCK 3 & 1.29 & 4.65 & 6.88\\
MOCK 4 & 1.55 & 9.33 & 15.45\\ 
MOCK 5 & 1.45 & 10.63 & 9.24\\ 
MOCK 6 & 0.38 & 6.21 & 8.96\\
MOCK 7 & 1.36 & 9.08 & 8.12\\
MOCK 8 & 0.18 & 6.91 & 4.27\\ 
MOCK 9 & 2.07 & 6.09 & 9.76\\ 
MOCK 10 & 1.62 & 4.40 & 11.91\\
MOCK 11 & 1.28  & 4.65 & 10.14\\ 
MOCK 12 & 2.65 & 7.97 & 11.25\\ 
MOCK 13 & 0.73 & 6.48 & 7.08\\
MOCK 14 & 0.36 & 7.30 & 16.44\\   
MOCK 15 & 0.98 & 4.36 & 8.47\\
MOCK 16 & 2.75 & 11.04 & 22.88\\ 
MOCK 17 & 0.30 & 5.96 & 7.67\\ 
MOCK 18 & 2.15 & 5.11 & 10.44\\
MOCK 19 & 1.59 & 8.02 & 10.99\\ 
MOCK 20 & 1.27 & 8.79 & 10.50\\
MOCK 21 & 2.77 & 10.57 & 11.06\\ 
MOCK 22 & 1.79 & 8.10 & 17.26\\   
\hline 
\end{tabular}
\caption{The mean of the sufficient statistics for the data and the mocks:  
$\bar{n_2}$ is the mean number of the $2$-connected cylinders,  
$\bar{n_1}$ is the mean number of the $1$-connected cylinders and  
$\bar{n_0}$ is the mean number of the $0$-connected cylinders.}
\label{tab:statdata}
\end{table} 

\begin{figure*} 
\centering
\resizebox{.33\hsize}{!}{\includegraphics*{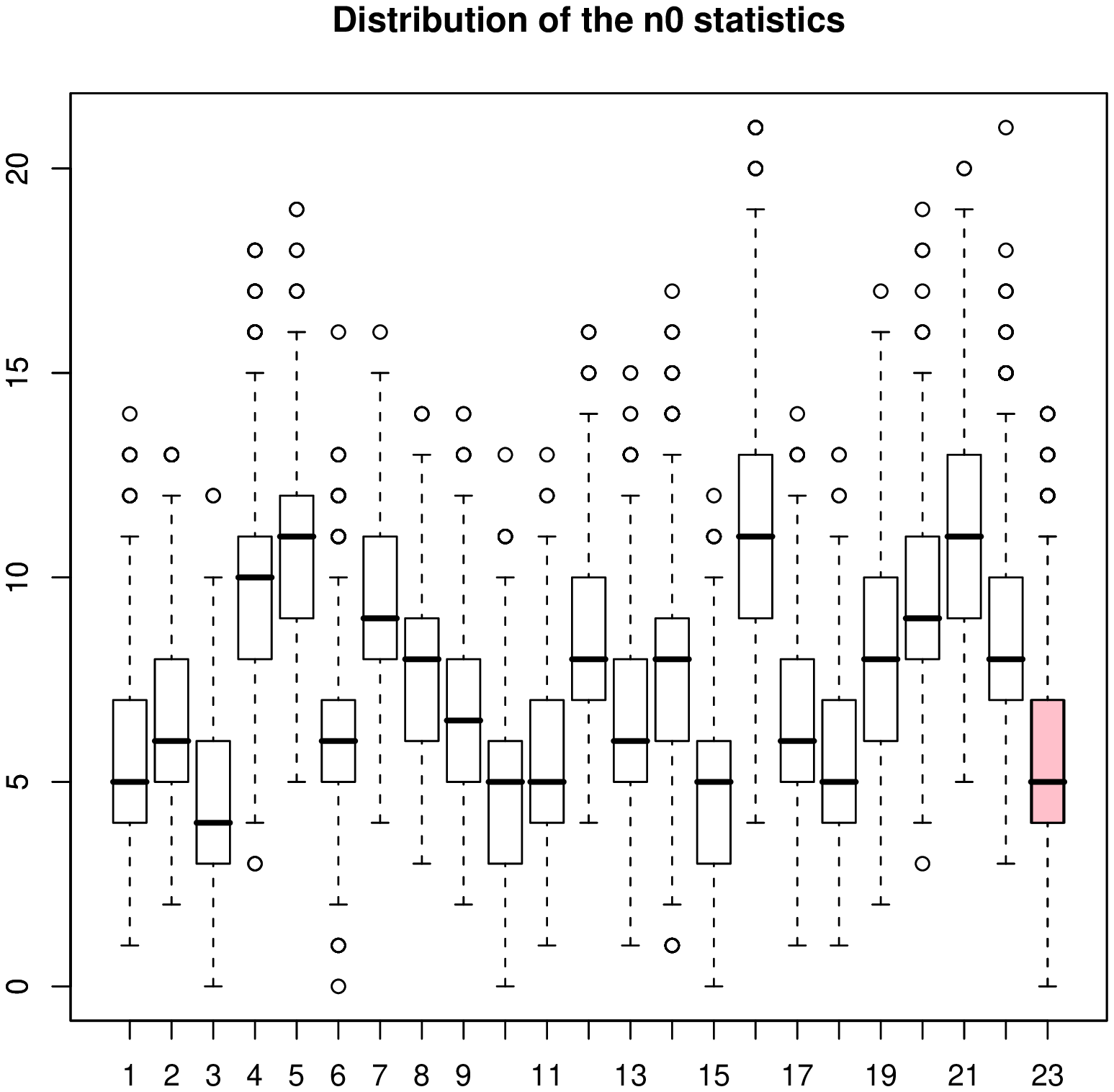}}
\resizebox{.33\hsize}{!}{\includegraphics*{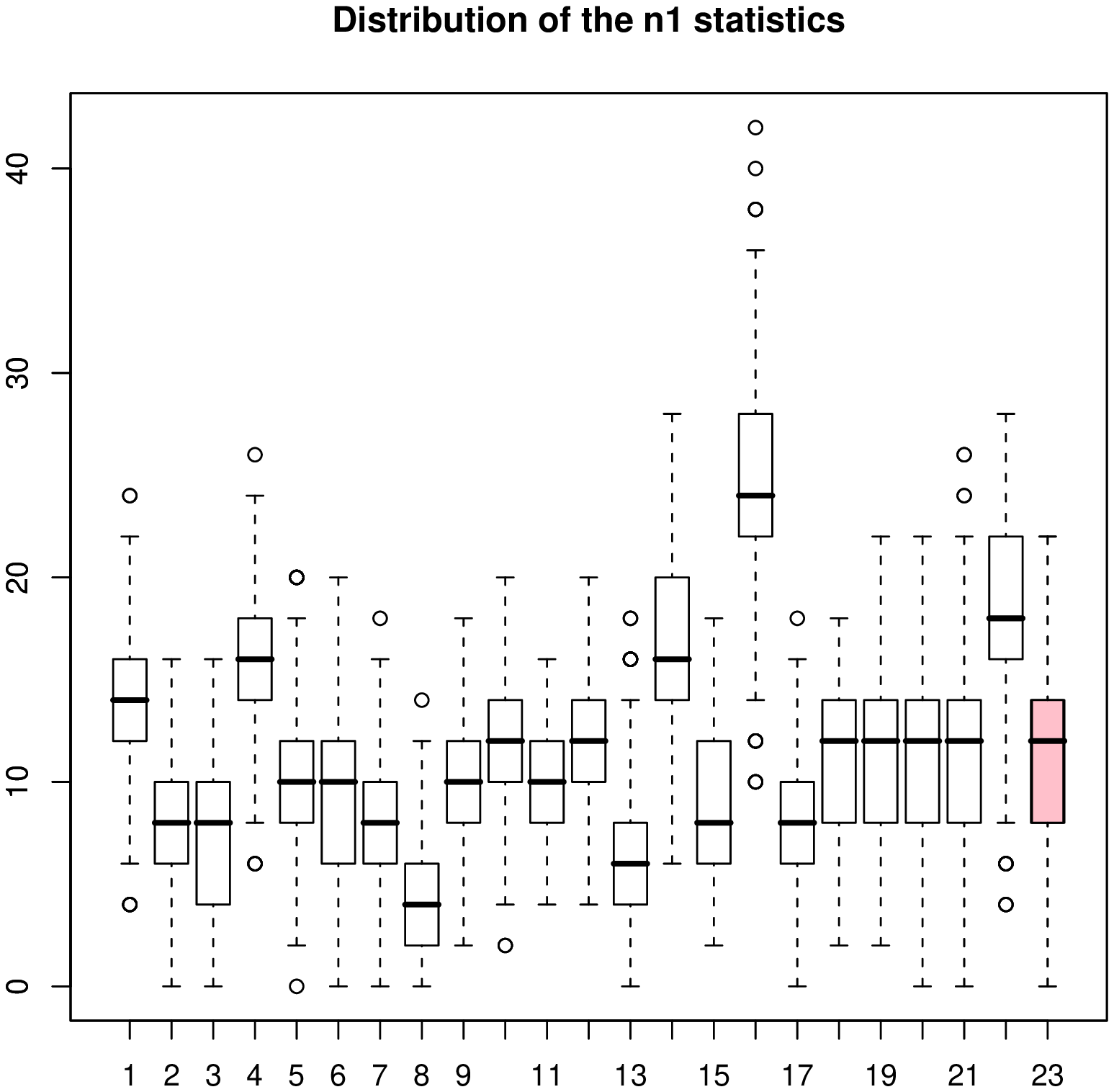}}
\resizebox{.33\hsize}{!}{\includegraphics*{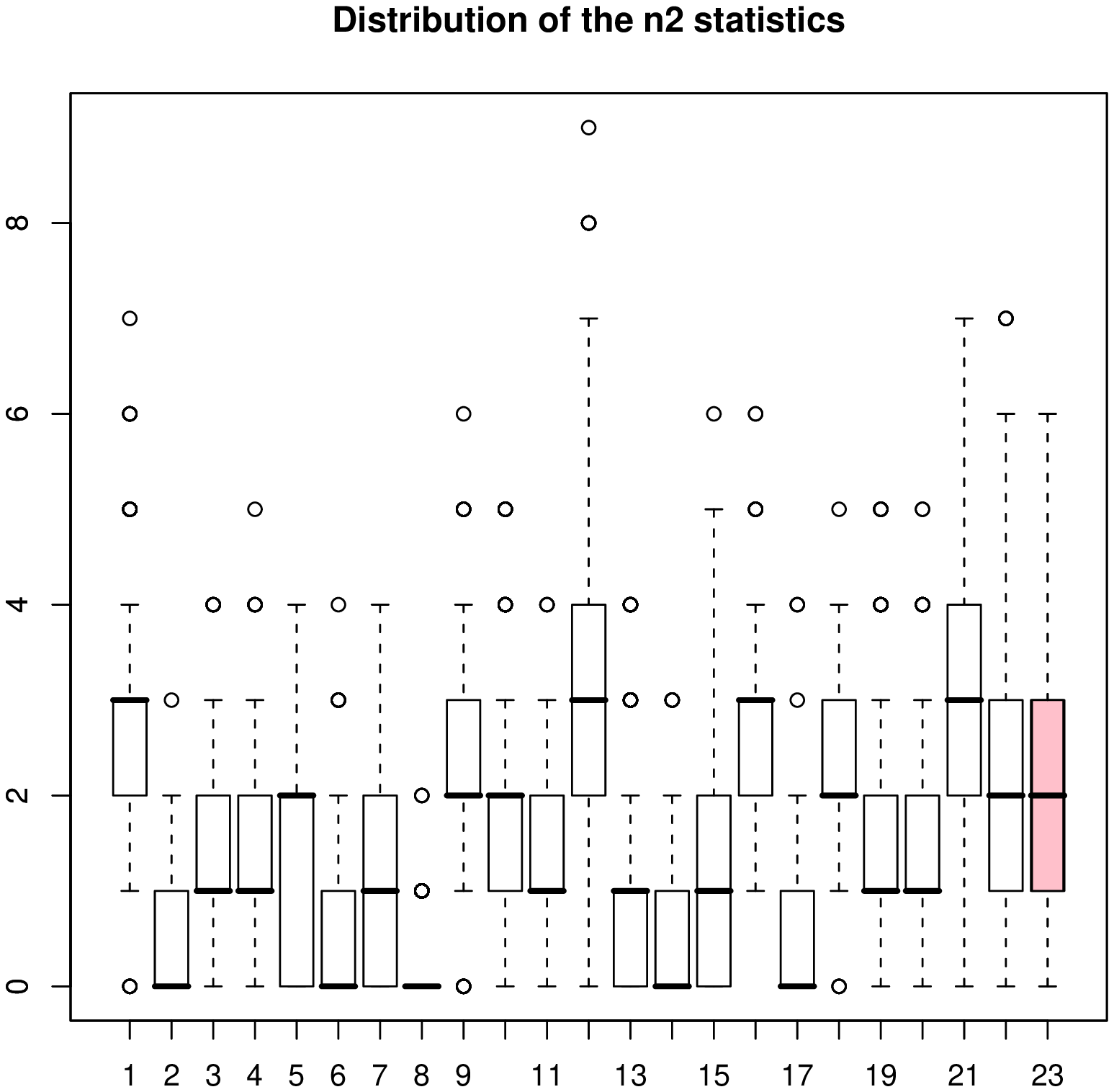}}\\
\caption{Comparison of the distributions of the sufficient statistics for
  the real data (dark boxplot) and the mock catalogues. 
  Box plots allow a simultaneous
  visual comparison in terms of the center, spread and overall range of the
  considered distributions: the middle line of the box corresponds to
  $q_{0.5}$, the empirical median of the distribution, whereas the
  bottom and the top of the box correpond to $q_{0.25}$ and $q_{0.75}$,
  the first and the third empirical quartile, respectively. The low extremal point
  of the vertical line (whiskers) of the box is given by
  $q_{0.25}-1.5\Delta q$, where the interquartile range $\Delta q=q_{0.75}-q_{0.25}$, 
  and the high extremal point is
  given by $q_{0.75}+1.5\Delta q$. There is a statistical
  rule-of-thumb stating that the values located outside the interval given
  by the whiskers may be outliers. These are shown by dots -- full dots
  indicate ``extreme'' outliers, more than $3\Delta q$ away from the $q_{0.25}$ or
  $q_{0.75}$, and open dots -- ``mild'' outliers, closer than $3\Delta q$ from these
  quartiles. As an example, for the Gaussian distribution the outlier region accounts
  for 0.7\% of the total probability.
}
\label{fig:boxplots}
\end{figure*}

The MH algorithm was run at a fixed temperature. This allows us a
quantitative comparison of the observed data and the mock catalogues
through the distributions of the model sufficient statistics. 

The box plots of the distributions of the sufficient statistics for
the mocks and the real data are shown in
Fig.~\ref{fig:boxplots}. The box plots for the mock catalogues are
indexed from $1$ to $22$, and this corresponds to the indexes of each
catalogue. The box plot corresponding to the real data is indexed
$23$ and it is coloured dark. 

The distributions of the $n_{0}$ statistics are compared
in the left panel in the Fig.~\ref{fig:boxplots}. We recall that the
$n_0$ statistics represents the number of isolated cylinders (no connections).
Thus, a large number of such cylinders tells us that the network
is rather fragmented. We see that only the mock catalogue 3
\mbox{exhibits} a less fragmented network than the real data.
A considerable number of mock catalogues has a filamentary
network that is much more fragmented than the data: the median for these
catalogues is clearly much higher than the median for the real
data. Nevertheless, there are some catalogues which have similar values as the data for
the median, and also a similar shape for the distributions. These mock
catalogues are $1,10,11,15$ and $18$. In order to see how similar
these catalogues are to the real data, we ran a Kolmogorov-Smirnov test.
The $p$-values for the mock catalogues $1$ and $18$
were $0.96$ and $0.13$, respectively. For the mock catalogues
$10,11$ and $15$ the obtained $p$-values were all lower than
$0.002$. Hence, we conclude that among all mock catalogues, there are only
two that are similar to the real data with respect to the
distribution of the $n_0$ statistics. A big
majority of the mock catalogues exhibits networks that are much more
fragmented than the one in the real data.

The right panel in Fig.~\ref{fig:boxplots} compares
the distributions of the $n_2$ statistics. The $n_2$ statistics
represents the number of cylinders in a configuration which are
connected at both its extremities. A configuration with a considerable
number of such cylinders forms a network made of rather long
filaments. So we see that
mock catalogues $12$ and $21$ exhibit a network with longer filaments
than in the data. The distribution of the $n_2$ statistics for the mock
catalogue $22$ appears to be very similar to the real data. The
$p$-value of the associated Kolmogorov-Smirnov test is $0.09$. 
To make a decision concerning the mock catalogues $1,9,16$ and
$18$, a \mbox{one-sided} Kolmogorov-Smirnov test was done, based on an
alternative hypothesis that the distribution of the $n_2$ statistics
for the real data may extend farther than the distributions for the
corresponding mock catalogs. Since the obtained $p$-values where all very
small, we can state that the filamentary network of these four mock catalogues
is made of shorter filaments than those in the real data. So, we
conclude that with a few exceptions the mock catalogues exhibit a network
made of filaments which are shorter than the ones in the real
data.

The distributions of the $n_1$ statistics are compared in the middle
panel of Fig.~\ref{fig:boxplots}. The $n_1$ statistics gives
the number of cylinders that are connected at only  
one extremity. Hence, a configuration with a large value of this
statistics is a network with a rather high density of the
filaments. Looking at all the three statistics together, we get a
rough idea of the topology of the network. For instance, if both the
values of $n_1$ and $n_2$ are high,
this indicates a network similar to a 
spaghetti plate or to a tree with long branches. Or the other way
around, if the $n_1$ and $n_0$ statistics are both high, the
network should be similar to a macaroni plate or
to a bush with short branches. This means that the mock
catalogues $1,4,14,16$ and $22$ produce a network which is
much more dense (or has more endpoints) than the one in
the real data catalogue. The box plots for the mock catalogs $18,19,20$
and $21$ are almost identical to the box plot for the
real data, showing that the $n_1$ distributions are similar.
This was confirmed by the 
Kolmogorov-Smirnov test. The mock catalogues $10$ and $12$ have the
same median as the data, while the distributions
are much more concentrated and symmetrical. The 
Kolomogorov-Smirnov test showed that these distributions differ significantly
from that for the real data. We conclude that
concerning the distributions of the $n_1$ statistics, four mock
catalogues have similar distributions to the data, and five others clearly have a much
more dense network, while the rest of the catalogues produce networks
that are clearly less dense or have fewer endpoints than
the network in the real data catalogue.

In summary, if for a single model characteristic the mock
catalogues may look similar to the data, taking into account all three
of them leads to a rather obvious difference between the mocks and the
observations. Generally, from a topological point of view, the networks
in the mock catalogues are more fragmented and contain shorter
filaments than in the data. As for the filament density,
the mock catalogues encompass the real data, with a large variance.

To see the influence of rescaling to the sufficient
statistics, we repeated the procedure with the rescaled cylinder for
the data and for the mocks 8 and 16. The data are given in
Table~\ref{tab:rescaled}.

\begin{table}
\centering
 \begin{tabular}{|l|r|r|r|} 
\hline 
& \multicolumn{3}{c|}{\strut Sufficient statistics}\\ 
\cline{2-4} \raisebox{1ex}[0pt]{Data sets}&\strut  $\bar{n_2}$ & $\bar{n_0}$  
& $\bar{n_1}$ \\ \hline 
\strut NGP250&11.31&32.76&56.15\\
2dF & 7.13  & 6.72 & 33.43\\ 
MOCK 8 & 1.53 & 9.57 & 12.64\\ 
MOCK 16 & 6.67 & 12.48 & 37.81\\ 
\hline 
\end{tabular}
\caption{The mean of the sufficient statistics for the data and the mocks, for the rescaled basic cylinder. The columns are the same as in the previous table.}
\label{tab:rescaled}
\end{table} 

Rescaling the basic cylinder improves the network, but not as much as expected -- the 
interaction parameters remain lower than those obtained for the NGP250 sample. 

To see if the filamentary network we find is really hidden in
the data, we uniformly re-distributed the points inside the domain
$K$. Now the points follow a binomial distribution that depends only
on the total number of points. For each (mock) data set this operation
was done $100$ times, obtaining $100$ point fields accordingly. For each
point field the method was launched during $50000$ iterations at fixed
$T=1.0$, while samples were picked up every $250$ iterations. The
model parameters were the same as previously described. The mean of
the sufficient statistics was then computed. The maximum values for
the all $100$ means for each data set are shown in
Table~\ref{tab:statbinomial}.

\begin{table} 
\centering
\begin{tabular}{|l|r|r|r|} 
\hline 
& \multicolumn{3}{c|}{\strut Sufficient statistics}\\ 
\cline{2-4} \raisebox{1ex}[0pt]{Binomial data sets}&\strut  $\max \bar{n_2}$ & $\max \bar{n_0}$ 
& $\max \bar{n_1}$ \\ \hline 
\strut MOCK 1 & 0  & 0.02 & 0\\ 
MOCK 2 & 0 & 0.015 & 0\\ 
MOCK 3 & 0 & 0.01 & 0\\
MOCK 5 & 0 & 0.015 & 0\\ 
MOCK 6 & 0 & 0.03 & 0\\
MOCK 7 & 0 & 0.02 & 0\\
MOCK 8 & 0 & 0.015 & 0\\
 
\hline 
\end{tabular}
\caption{The maximum of mean of the sufficient statistics over binomial 
fields generated for some mock catalogues (the same number of points):  $\max\bar{n_2}$ is the maximum mean number of the $2$-connected cylinders,  $\max\bar{n_1}$ is the maximum mean number of the $1$-connected cylinders and  $\max\bar{n_0}$ is the maximum mean number of the $0$-connected cylinders. 
}
\label{tab:statbinomial} 
\end{table}

As we see, the algorithm does not find any
connected cylinders for a random distribution, both the numbers of the $1$-connected and
$2$-connected cylinders are strictly zero.  Only in a few cases the
data allow us to place a single cylinder. Thus, the filaments our
algorithm discovers in galaxy surveys and in mock catalogues are real,
they are hidden in the data and are not the result of a lucky choice
of the model parameters.

\section{Discussion, conclusions and perspectives}

In a previous paper \citep{StoiMartSaar07} we developed a new approach
to locate and characterise filaments hidden in three-dimensional point
fields. We applied it to a galaxy catalogue (2dFGRS), found the
filaments and described their properties by the sufficient statistics
(interaction parameters) of our model.

As there are numerical models (mocks) that are carefully constructed
to mimic all local properties of the 2dFGRS, we were interested to see whether
these models also have global properties similar to the observed
data. An obvious test for that is to find and compare the filamentary
networks in the data and in the mocks. We did that, using fixed shape
parameters for the basic building blocks for the filaments, and fixed
interaction potentials. These priors had led to good results before.

In order to strictly compare the observed catalogue and the mocks, we
had to work with constant-density samples (volume-limited
catalogues). This inevitably led to a smaller spatial density, and the
filament networks we recovered were not as good as those found in the
previous paper. Rescaling the basic cylinder helped, but not as much
as expected.
 
As all the mock samples are selected from a single large-volume
simulation, they share the same realisation of the initial density
and velocity fields. The large-scale properties of the density field and
its filamentarity should be similar in all the mocks. The volumes of
the mocks are sufficiently high to suppress cosmic noise at the filament
scales, and the dimensions of the bricks are large, too, except for the third
dimension, the thickness of the brick. Our bricks are very thin, with a height
of only $31.1h^{-1}$Mpc). This can cause a selection of different pieces
of dark matter filaments and consequently a broad variance in the filamentarity of
the density field. 

The biasing scheme will also influence the properties of mock filaments.
As the particle mass in
the simulation was $2.2\cdot10^{12}\mbox{M}_{\sun}$, galaxies had to be identified
with individual mass points, and this makes the biasing scheme pretty random
(compared with later scenarios where galaxies have been built inside dark
matter subhaloes). Another source of randomness is the random assignment of
galaxy luminosities that excludes reproducing the well-known luminosity-density relation.
As we saw, the filaments in a typical mock are shorter, and that can be explained
by the 'randomisation' of galaxy chains.

There are several new results in our paper:
\begin{enumerate}
\item The filamentarity of the real galaxy catalogue, as described by
  the sufficient statistics of our model (the interaction parameters),
  lies within the range covered by the mocks. But the model
  filaments are, in general, much shorter and do not form an extended network.
\item The filamentarity of the mocks themselves differs much.
  This may be caused both by the specific geometry (thin slice) of the sample 
  volume and by the biasing scheme used to populate the mocks with galaxies.
\item Finally, we compared our catalogues with the random (binomial)
  catalogues with the same number of data points and found that these
  do not exhibit any filamentarity at all. This proves that the
  filaments we find exist in the data.
\end{enumerate}

Our method
does not yield an estimate of the precision of the
detection. This is an important and far from a trivial \mbox{problem}. For
instance, even in the ideal situation of an entirely supervised
detection of these filaments (made by hand by a human specialist), we
may wonder how the obtained result should be validated? Another very
important difficulty is related to the fact that there
exists no precise (mathematical) definition of what a galactic filament really is.

The definition we propose, ``something complicated made of connected
small cylinders containing galaxies that are more
concentrated in a cylinder than outside it'', can be clearly
improved in order to allow a better local fit for a
cylinder. Still, although quite simple, this definition allows us a general
treatement using marked point processes. Very recent work
in marked point process literature presents methodological ideas
leading to statistical model
validation \citep{BaddTurnMollHaze05,BaddMollPake08}. This gives 
hope and perspective to incorporate these ideas into our method. This
perspective is important because it allows us a global appreciation of
the result.

The detection test on realisations of binomial point processes shows
that whenever filaments are not present in the data, the proposed
method does not detect filaments. This also means that the
detected filaments in the data are ``true filaments'' (in the sense of
our definition) and not a ``random alignment of points'' (false
alarms) that may occur by chance even in a binomial point process. In
that sense, together with the topological information given by the
sufficient statistics, our model is a good tool for describing the
network. The strong point of this approach is that it allows simultaneous
morphological description and statistical inference. Another important
advantage of using a marked point process based methodology is that it
allows for the evolution of the definition of the objects forming the
pattern we are looking for.

One of the messages this paper communicates is that looking at two
different families of data sets with the same statistical tool, we get
rather different results from a statistical point of view. Therefore,
we can safely conclude that the two families of data sets are different.

There are many ways to improve on the work we have done so far. We
have seen above that it is difficult to find the scale (lengths) of
the filaments for our model; this problem has to be solved. Second, we
have used fixed parameters for the data term (cylinder sizes); these
should be found from the data. Third, the filament network seems to be
hierarchical, with filaments of different widths and sizes; a good
model should include this. 
Fourth, parameter estimation and detection
validation should be also included; the uniform law does not allow
the characterisation of the model parameters distribution and for the
moment we cannot say that the detected filamentary pattern is
correctly detected; the only statistical statement that we can do is
that this pattern is hidden in the data and we have some good ideas
about where it can be found, but we do not give any precise measure
about it.

Also, it would be good if our model could be extended to describe
inhomogeneous point processes -- magnitude-limited catalogues that
have much more galaxies and where the filaments can be traced much
better. The first rescaling attempt we made in this paper could be
a step in this direction, but as we saw, it is not perfect. And, as
usual in astronomy -- we would understand nature much better if we
had more data. The more galaxies we see at a given location, the
better we can trace their large-scale structure.

The Bayesian framework and the theory of marked point process allow
the mathematical formulation for filamentary pattern detection
methodologies introducing the previously mentioned improvements
(inhomogeneity, different size of objects, \mbox{parameter} estimation).
The numerical implementation and the construction of these
improvements in harmony with the astronomical observations and
theoretical knowledge are open and challenging problems.

\begin{acknowledgements}
First, we thank our referee for detailed and constructive criticism and
suggestions.
This work has been supported by the University of Valencia through a
visiting professorship for Enn Saar, by the Spanish Ministerio de
Ciencia e Innovaci\'on CONSOLIDER projects AYA2006-14056 and
CSD2007-00060, including FEDER contributions, by the Generalitat 
Valenciana  project of excellence PROMETEO/2009/064, by the Estonian Ministry
of Education and Science, research project SF0060067s08, and by the
Estonian Science Foundation grant 8005. We thank D. Croton for the
observational data (the 2dFGRS volume-limited catalogues) and the mock
catalogues. The authors are also grateful to G. Castellan for helpful
discussions concerning statistical data analysis.

Three-dimensional visualisation in this paper was conducted with the
S2PLOT programming library \citep{Barnes06}. We thank the S2PLOT team
for superb work.
\end{acknowledgements}

\bibliographystyle{aa}
\bibliography{12823}

\end{document}